\documentclass[draft]{agujournal2019}
\usepackage{url} %this package should fix any errors with URLs in refs.
\usepackage{soul}
\draftfalse

\journalname{JGR: Atmospheres}

\begin{document}

%% ------------------------------------------------------------------------ %%
%  Title
%
% (A title should be specific, informative, and brief. Use
% abbreviations only if they are defined in the abstract. Titles that
% start with general keywords then specific terms are optimized in
% searches)
%
%% ------------------------------------------------------------------------ %%

% Example: \title{This is a test title}

\title{Climatologies of Various OH Lines From About 90,000 X-shooter 
Spectra}

%% ------------------------------------------------------------------------ %%
%
%  AUTHORS AND AFFILIATIONS
%
%% ------------------------------------------------------------------------ %%

% Authors are individuals who have significantly contributed to the
% research and preparation of the article. Group authors are allowed, if
% each author in the group is separately identified in an appendix.)

% List authors by first name or initial followed by last name and
% separated by commas. Use \affil{} to number affiliations, and
% \thanks{} for author notes.
% Additional author notes should be indicated with \thanks{} (for
% example, for current addresses).

% Example: \authors{A. B. Author\affil{1}\thanks{Current address, Antartica}, B. C. Author\affil{2,3}, and D. E.
% Author\affil{3,4}\thanks{Also funded by Monsanto.}}

\authors{S. Noll\affil{1,2}, C. Schmidt\affil{2}, W. Kausch\affil{3}, 
  M. Bittner\affil{2,1}, S. Kimeswenger\affil{3,4}}

% \affiliation{1}{First Affiliation}
% \affiliation{2}{Second Affiliation}
% \affiliation{3}{Third Affiliation}
% \affiliation{4}{Fourth Affiliation}

\affiliation{1}{Institut f\"ur Physik, Universit\"at Augsburg, Augsburg,
  Germany}
\affiliation{2}{Deutsches Fernerkundungsdatenzentrum, Deutsches Zentrum f\"ur 
  Luft- und Raumfahrt, Oberpfaffenhofen, Germany}
\affiliation{3}{Institut f\"ur Astro- und Teilchenphysik, Universit\"at
  Innsbruck, Innsbruck, Austria}
\affiliation{4}{Instituto de Astronom\'ia, Universidad Cat\'olica del Norte,
  Antofagasta, Chile}

%(repeat as many times as is necessary)

%% Corresponding Author:
% Corresponding author mailing address and e-mail address:

% (include name and email addresses of the corresponding author.  More
% than one corresponding author is allowed in this LaTeX file and for
% publication; but only one corresponding author is allowed in our
% editorial system.)

% Example: \correspondingauthor{First and Last Name}{email@address.edu}

\correspondingauthor{Stefan Noll}{stefan.noll@dlr.de}

%% Keypoints, final entry on title page.

%  List up to three key points (at least one is required)
%  Key Points summarize the main points and conclusions of the article
%  Each must be 140 characters or fewer with no special characters or punctuation and must be complete sentences

% Example:
% \begin{keypoints}
% \item	List up to three key points (at least one is required)
% \item	Key Points summarize the main points and conclusions of the article
% \item	Each must be 140 characters or fewer with no special characters or punctuation and must be complete sentences
% \end{keypoints}

\begin{keypoints}
\item Climatologies of intensity, solar cycle effect, and residual variability
of 298 OH lines were derived from 10 years of X-shooter data
\item The strongest variations are found for intermediate rotational energies 
where cold and hot OH populations show similar contributions
\item Tides cause a local time shift of the climatological patterns depending 
on the effective line emission height 
\end{keypoints}

%% ------------------------------------------------------------------------ %%
%
%  ABSTRACT and PLAIN LANGUAGE SUMMARY
%
% A good Abstract will begin with a short description of the problem
% being addressed, briefly describe the new data or analyses, then
% briefly states the main conclusion(s) and how they are supported and
% uncertainties.

% The Plain Language Summary should be written for a broad audience,
% including journalists and the science-interested public, that will not have 
% a background in your field.
%
% A Plain Language Summary is required in GRL, JGR: Planets, JGR: Biogeosciences,
% JGR: Oceans, G-Cubed, Reviews of Geophysics, and JAMES.
% see http://sharingscience.agu.org/creating-plain-language-summary/)
%
%% ------------------------------------------------------------------------ %%

%% \begin{abstract} starts the second page

\begin{abstract}
The nocturnal mesopause region of the Earth's atmosphere radiates 
chemiluminescent emission from various roto-vibrational bands of hydroxyl 
(OH), which is therefore a good tracer of the chemistry and dynamics at the 
emission altitudes. Intensity variations can, e.g., be caused by the 
general circulation, gravity waves, tides, planetary waves, and the solar 
activity. While the basic OH response to the different dynamical influences
has been studied quite frequently, detailed comparisons of the various 
individual lines are still rare. Such studies can improve our understanding 
of the OH-related variations as each line shows a different emission profile. 
We have therefore used about 90,000 spectra of the X-shooter spectrograph of 
the Very Large Telescope at Cerro Paranal in Chile in order to study 10 years 
of variations of 298 OH lines. The analysis focuses on climatologies of 
intensity, solar cycle effect, and residual variability (especially with 
respect to time scales of hours and about 2 days) for day of year and 
local time. For a better understanding of the resulting variability patterns 
and the line-specific differences, we applied decomposition techniques, 
studied the variability depending on time scale, and calculated 
correlations. As a result, the mixing of thermalized and nonthermalized OH 
level populations clearly influences the amplitude of the variations. 
Moreover, the local times of the variability features shift depending on the 
effective line emission height, which can mainly be explained by the 
propagation of the migrating diurnal tide. This behavior also contributes to 
remarkable differences in the effective solar cycle effect.  
\end{abstract}

\section*{Plain Language Summary}
Emission from various lines of the hydroxyl (OH) molecule is an important 
contribution to the Earth's nighttime radiation in the near-infrared. The
emission mostly originates from altitudes between 80 and 100\,km and is 
therefore a good tracer of the chemistry and dynamics at these altitudes. 
OH intensity variations can be caused by changes in the atmospheric 
conditions and passing waves with different time scales. In order to better 
understand the origin of these variations and their impact on the OH 
emission, we studied the variability of 298 OH lines measured in 10 years of 
data from the X-shooter spectrograph at Cerro Paranal in Chile. The analysis
focused on average variations with respect to local time and day of year, 
i.e. climatologies. As the lines show different vertical emission 
distributions, this study also provides height-dependent information. The 
climatologies for intensity, the response to the solar activity cycle of 11 
years, and the residual variability (dominated by waves with time scales of a 
few hours and about 2 days) revealed remarkable patterns which depend on the 
OH excitation level. The features can partly be explained by the impact of 
solar tides, particularly with a period of 24\,h.

%% ------------------------------------------------------------------------ %%
%
%  TEXT
%
%% ------------------------------------------------------------------------ %%

%%% Suggested section heads:
% \section{Introduction}
%
% The main text should start with an introduction. Except for short
% manuscripts (such as comments and replies), the text should be divided
% into sections, each with its own heading.

% Headings should be sentence fragments and do not begin with a
% lowercase letter or number. Examples of good headings are:

% \section{Materials and Methods}
% Here is text on Materials and Methods.
%
% \subsection{A descriptive heading about methods}
% More about Methods.
%
% \section{Data} (Or section title might be a descriptive heading about data)
%
% \section{Results} (Or section title might be a descriptive heading about the
% results)
%
% \section{Conclusions}

\section{Introduction}\label{sec:intro}
%Text here ===>>>

Chemiluminiscent emission of the hydroxyl (OH) radical dominates the 
nocturnal radiation of the Earth's atmosphere in the near-infrared 
wavelength regime. Various roto-vibrational bands of the electronic ground 
state contribute to the emission spectrum \cite<e.g.,>{noll15,rousselot00}. 
The radiation originates in the mesopause region between 80 and 100\,km 
\cite<e.g.,>{baker88,noll22} and is mostly related to the production of OH 
with relatively high vibrational excitation (up to a vibrational level 
$v = 9$) by the reaction of atomic hydrogen and ozone \cite{bates50} and the 
subsequent relaxation processes. Apart from the emission of photons, 
collisions with different constituents of the atmosphere contribute to this 
redistribution of the level populations. In the end, the population of each 
$v$ can be described by a cold, fully thermalized and a hot, nonthermalized 
component \cite{kalogerakis18,noll20,oliva15}. The latter dominates the 
populations of levels with high rotational quantum numbers $N$.
 
The OH emission layer with a typical full width at half maximum of about 
8\,km \cite{baker88} can be affected by perturbations in pressure, 
temperature, and the distribution of the atmospheric constituents on 
different time scales as such changes alter the production and thermalization
of OH molecules. In particular, atomic oxygen matters. This radical is
required for the production of ozone and plays an important role in the
vibrational relaxation and destruction of OH 
\cite<e.g.,>{adler97,dodd94,noll18b,savigny12}. It shows a strong response to
vertical transport since its concentration steeply declines in the lower part 
of the OH emission region \cite<e.g.,>{smith10}, where the buildup of a 
reservoir by photolysis of molecular oxygen at daytime is less efficient and 
the consumption of atomic oxygen at nighttime is fast under undisturbed 
conditions \cite{marsh06}. An important source of perturbation are the 
globally acting solar tides, especially the westward migrating diurnal and 
semidiurnal tides \cite<e.g.,>{smith12}, which propagate from the 
troposphere/stratosphere into the mesopause region \cite<e.g.,>{hagan95}. The 
related changes in the vertical pressure, temperature, and chemical 
composition profiles can significantly alter the nocturnal trend in OH 
emission, i.e. emission increases are also possible 
\cite<e.g.,>{marsh06,takahashi98,zhang01}. Intra-annual changes in the tidal 
amplitudes also contribute to the observed seasonal variability of OH 
emission with maximum intensities around the equinoxes at low latitudes
\cite<e.g.,>{gao10,shepherd06,takahashi95}. Another source of perturbations
of the mesopause region are gravity waves, which have periods from minutes to 
hours and act on a regional scale \cite<e.g.,>{fritts03}. Individual gravity 
waves are sporadic but the general activity of such waves with respect to OH
emission shows a seasonal pattern 
\cite<e.g.,>{hannawald19,kim10,lopez20,reisin04,sedlak20}. Increased activity
tends to be observed around solstices or in winter depending on the wave
period and the latitude. The intra-annual variations are related to the 
weather conditions in the troposphere (the dominating source region for 
primary waves) and the wind speeds and directions up to the mesosphere. The 
latter rule the efficiency of the blocking of the vertical wave propagation
and the related possible generation of secondary waves. OH emission is also 
influenced by the globally acting planetary waves with periods of the order 
of days \cite<e.g.,>{lopez09,pedatella12} and the seasonal changes of the 
residual meridional circulation \cite<e.g.,>{marsh06}. Changes in OH 
nightglow on time scales of the order of years are particularly caused by the 
solar activity cycle of about 11 years \cite<e.g.,>{gao16,noll17}, which 
leads to a significant variation of hard ultraviolet photons that can, e.g.,
destroy molecular oxygen \cite<e.g.,>{marsh07}. 

In conclusion, the sensitivity of OH lines to the various sources of 
variation makes them valuable for the study of the dynamics in the mesopause 
region. However, the investigations are often based on a few bright lines or 
unresolved broad-band data due to instrumental limitations. This constitutes
a loss of information. As the radiative lifetimes and rate coefficients for
collisions depend on the specific OH energy level, the vertical emission
distributions deviate for OH lines with different upper levels
\cite<e.g.,>{dodd94,noll18b,savigny12}. Consequently, the response of OH 
emission to perturbations depends on the selected line. The resulting 
differences can therefore provide additional information on the vertical
component of the dynamics as well as the OH-related chemistry. At least for
the integrated intensities of the Q branches of OH(3-1) and OH(4-2), this 
was demonstrated by \citeA{schmidt18}, who estimated vertical wavelengths of 
gravity waves. Studies of large sets of individual OH lines are rare, 
particularly with respect to variations. With a few thousand spectra in 
maximum \cite{cosby07,hart19a,noll15,noll17}, only a rough characterization 
of the dynamics was possible. Some differences in the nocturnal, seasonal, 
and long-term variations for lines with different upper vibrational levels 
$v^{\prime}$ were identified. A fraction of these variations might be 
explained by differences in the thermalization of the involved level 
populations as the observed changes in the rotational and vibrational 
temperatures, i.e. ratios of lines with different upper levels $N^{\prime}$ 
and $v^{\prime}$, indicate. The studies did not involve weak lines with high 
$N^{\prime}$, which would be crucial for a better understanding of the 
variation of the level populations on different time scales especially with 
respect to the hot component.        

In a previous study \cite{noll22}, we analyzed 298 OH lines with a wide range 
of $v^{\prime}$ and $N^{\prime}$. The intensities were measured in spectra of 
the X-shooter spectrograph \cite{vernet11} of the Very Large Telescope (VLT) 
of the European Southern Observatory (ESO) at Cerro Paranal in Chile 
(24.6$^{\circ}$\,S, 70.4$^{\circ}$\,W). In a time interval of eight nights in 
January/February 2017 (and also seven nights in January 2019), the data set
allowed us to perform a detailed study of the propagation of a very strong
(and a moderate) quasi-2-day wave (Q2DW), which is the most remarkable
planetary wave at low southern latitudes \cite<e.g.,>{ern13,gu19,tunbridge11}.
It is characterized by a lifetime of only a few weeks usually in summer and a
westward moving pattern with a zonal wavenumber of 3 in the southern
hemisphere. Our fits of the wave properties resulted in a most likely period
of 44\,h for both events, a strong dependence of the wave amplitude on local
time in 2017 mainly due to the interaction of the Q2DW with solar tides, and
maximum amplitudes for lines with intermediate $N^{\prime}$ in both years. As
the latter show similar contributions of cold and hot populations
\cite{noll20}, the increased intensity variability can be explained by
variations in the ambient temperature that significantly affect the population
mixing. In combination with emission profiles of the two OH channels of the
Sounding of the Atmosphere using Broadband Emission Radiometry (SABER)
instrument onboard the Thermosphere Ionosphere Mesosphere Energetics Dynamics
(TIMED) satellite \cite{russell99}, we linked wave phases and emission heights
for the wave in 2017 thanks to a nearly linear relation and significant phase
differences. With a vertical wavelength of about 32\,km, we finally derived
average centroid emission heights between 86 and 94\,km. The emission
altitudes increase with $v^{\prime}$ and $N^{\prime}$.

These results demonstrate the potential of parallel investigations of the 
dynamics of various OH lines. In this study, we extend the analysis of the 
same 298 lines by using the entire X-shooter data set discussed by 
\citeA{noll22} of about 90,0000 near-infrared spectra covering a period of 10 
years between October 2009 and September 2019. The resulting wealth of data
allowed us to study climatologies for local time and day of year for the 
intensity, solar cycle effect, and residual variability in order to 
characterize the impact of the OH level on the intensity variation for 
pertubations with different time scales. The contents of the paper are as 
follows. First, we will briefly describe the data set 
(section~\ref{sec:data}). Then, we will introduce our approach to calculate 
the climatologies and our decomposition techniques for a detailed analysis 
(section~\ref{sec:methods}). The results for the different investigated 
properties will be discussed in section~\ref{sec:results}. Finally, we will 
draw our conclusions (section~\ref{sec:conclusions}).

\section{Data}\label{sec:data}

We used VLT/X-shooter spectra in the near-infrared between 1.0 and 
2.5\,$\mu$m \cite{vernet11} taken in a time interval of 10 years. In the 
following, we give a brief overview of the related data processing and data 
selection. More details are provided by \citeA{noll22}.

The raw medium-resolution echelle spectra of astronomical targets originate 
from the ESO Science Archive Facility and were first processed with version 
v2.6.8 of the official reduction pipeline \cite{modigliani10} and 
preprocessed calibration data. The resulting wavelength-calibrated 
two-dimensional (2D) spectra were then further treated by averaging them 
along the spatial direction (projected slit length of 11$^{\prime\prime}$), 
correcting them for systematic biases in the pipeline-based separation of 
sky and astronomical target, and performing an absolute flux calibration 
based on 10 master response curves with valid time intervals between 9 and 15
months that were derived from X-shooter spectra of the spectrophotometric 
standard stars EG\,274 and LTT\,3218 \cite{moehler14}. For the predominating
clear sky conditions, the relative flux uncertainties are expected to be of 
the order of 2 to 3\% for wavelengths up to 2.1\,$\mu$m. The accuracy of the
absolute fluxes is lower due to uncertainties in the reference spectral 
energy distributions of the standard stars. 

In the final spectra, the continuum was determined by means of the lowest 
quintile of the intensities in pixel ranges that depended on the
corresponding line density and slope of the continuum. After the subtraction
of the continuum, line intensities were measured in wavelength ranges which 
depended on the variable width of the entrance slit of the spectrograph and 
the separation of the $\Lambda$ doublet components of each OH line as taken 
from \citeA{brooke16}. The measured line intensities were corrected for their 
dependence on the zenith angle due to projection effects assuming a reference 
altitude of 87\,km. Moreover, we considered the partial absorption of line 
emission by molecules especially in the lower atmosphere. The line-specific 
absorption was modeled for Doppler-broadened OH lines assuming a temperature 
of 190\,K and by means of the Line-By-Line Radiative Transfer Model 
\cite<LBLRTM,>{clough05}, which involved typical atmospheric profiles for 
Cerro Paranal \cite{noll12}. Apart from the line, the correction depended on 
the zenith angle and the highly variable amount of water vapor. The latter 
was estimated based on pairs of OH lines with very different absorption. The 
relations were calibrated using data of the Low Humidity And Temperature 
PROfiler (L-HATPRO) microwave radiometer at Cerro Paranal \cite{kerber12}. 
For this study, we used the same 298 OH lines as selected for the analysis of 
the Q2DW in 2017 by \citeA{noll22}. In this way, we can directly compare both 
investigations. Moreover, the results suggest that an additional optimization 
of the line set for the studied climatological properties would not 
significantly improve the quality of the analysis. Selection criteria such as 
high atmospheric transmission, negligible line blending, and smooth 
underlying continuum work independently of the data sample and the analyzed 
property.

\begin{figure}
\noindent\includegraphics[width=\textwidth]{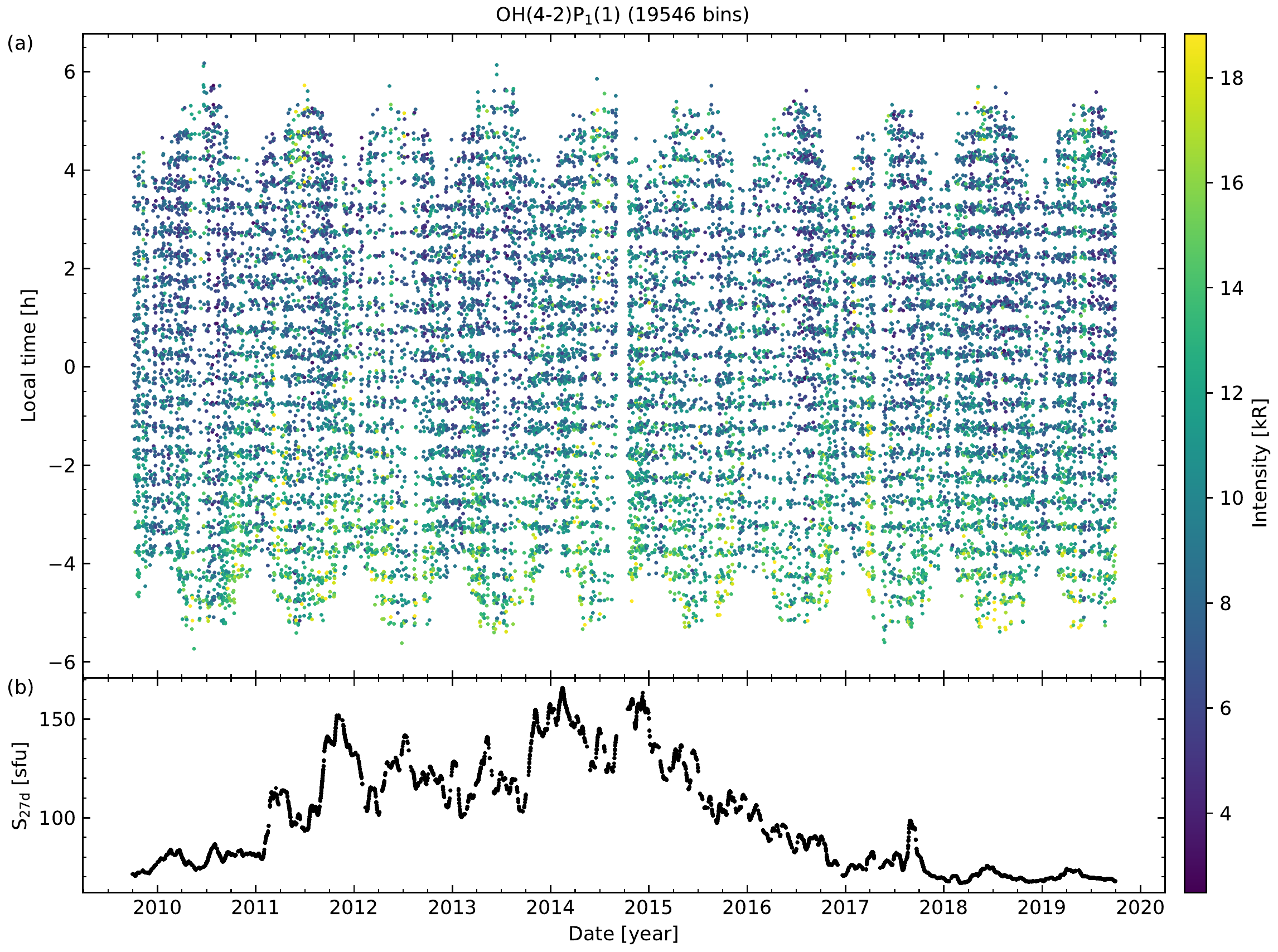}
\caption{Sampling of date in years and local time (LT) in hours for the 19,546
30\,min bins used for OH(4-2)P$_1$(1) (a). The given times of each LT interval
differ due to the different central times and exposure lengths of the
contributing individual observations. (a) also provides the mean intensities
of the bins in kilorayleigh (kR). For a better visibility of the variations,
the upper limit of the color scale was fixed to 3 standard deviations above
the mean intensity. This cut only affected 132 bins. (b) shows the sampling of
the solar radio flux at 10.7\,cm averaged for 27 days, $S_\mathrm{27d}$, in
solar flux units (sfu). The distribution indicates that the most of solar
cycle 24 is covered.}
\label{fig:sampling}
\end{figure}

As the individual spectra show strong variations in the exposure time,
spectral resolution, absorption by water vapor, zenith angle, and residual
contribution of the astronomical target, the size of the useful data set
depends on OH line properties such as intensity and wavelength. Based on a
$\sigma$-clipping approach for outlier detection with respect to continuum,
intensity, and intensity uncertainty, the final line-specific samples comprise
between 61,458 and 88,315 data points with a mean of 82,836 for the 298 lines.
In order to further improve the quality, we averaged the intensities for
consecutive 30\,min intervals and only kept those intervals with a minimum
summed exposure time of 10\,min. This approach significantly reduced the
variation in the size of the data set. For the 268 OH lines with wavelengths
up to 2.1\,$\mu$m, the resulting number of bins is between 18,936 and 19,570
(mean of 19,480 and relative standard deviation of 0.65\%). The corresponding
data coverage with respect to date and local time is illustrated in
Figure~\ref{fig:sampling}a for an example line. In general, there is a
relatively smooth coverage as 63\% of the nights are covered with an average
number of 8.5 bins. Data gaps longer than a week are rare (maximum of 41
consecutive nights in 2014). As bad weather is a minor issue at Cerro Paranal
\cite{holzloehner21,kerber16}, technical reasons are more common (especially
telescope sharing by different instruments). In the case of the 30 lines at
longer wavelengths (mainly belonging to OH(9-7)), the bin numbers range from
16,926 to 17,019 (mean of 17,001 and relative standard deviation of 0.10\%).
Hence, the only noteworthy differences are related to the wavelength regime.
Spectra taken with a so-called $K$-blocking filter for straylight reduction
\cite{vernet11} cannot be used beyond 2.1\,$\mu$m. Nevertheless, the decrease
of the sample size at long wavelengths is only about 13\%. Hence, the
resulting climatologies should still be sufficiently consistent. Comparisons 
for $v^{\prime} = 9$ lines from different bands did not show clear 
discrepancies.

\section{Methods}\label{sec:methods}

\subsection{Calculation of Climatologies}\label{sec:calcclim}

\begin{figure}
\noindent\includegraphics[width=\textwidth]{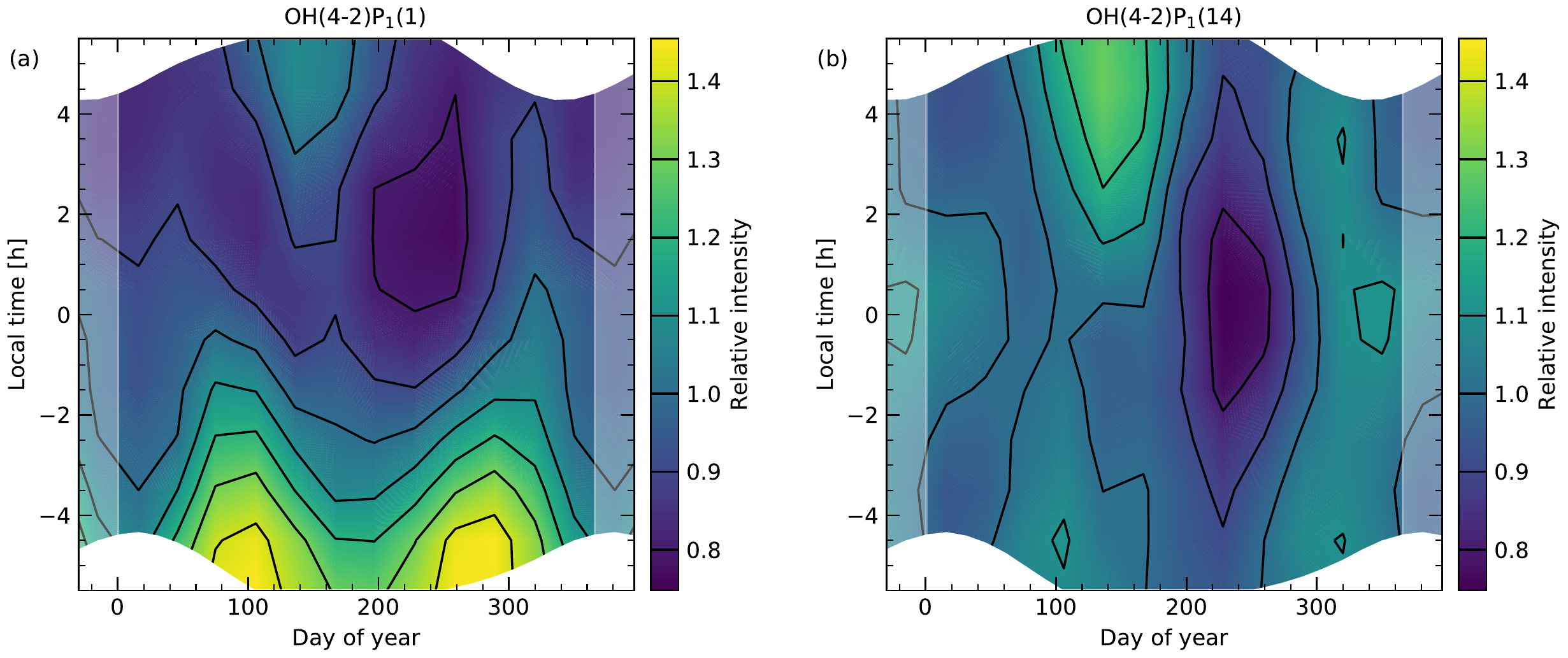}
\caption{Climatologies of intensity relative to mean as a function of local 
time (with a resolution of 1\,h) and day of year (with a resolution of 1
month) for OH(4-2)P$_1$(1) (a) and OH(4-2)P$_1$(14) (b). The climatologies 
are representative of a solar radio flux of 100\,sfu. The colored contours 
are only provided for dates and times with solar zenith angles larger than 
100$^{\circ}$. The seasonal variation is partly repeated (marked by lighter
colors) for a better representation around the turn of the year.}
\label{fig:i_2lin}
\end{figure}

In this study, we focus on 2D climatologies depending on local time (LT) and 
day of year (e.g. Figure~\ref{fig:i_2lin}). The local time is defined as mean 
solar time for the longitude of Cerro Paranal (70.4$^{\circ}$\,W). Each 
climatology consists of a grid of $12 \times 12$ data points with centers in 
the middle of the nighttime hours (from 18:30 to 05:30 LT) and the middle of 
the months. Each grid point represents the average of a property for a 
selection of close data bins, which were derived as discussed in 
section~\ref{sec:data} and are representative of the average of the 
central times of the considered individual exposures weighted by the exposure 
time (see Figure~\ref{fig:sampling}a). The maximum relative distance to a grid
point was usually 1.0, which corresponds to time differences of 1\,h and 1
month of average length, respectively. Hence, a bin can contribute to several
adjacent grid points. The climatologies are smoothed. Smaller selection radii
would lead to a better time resolution but worse statistics. We required a
minimum number of selected bins of 400. Even if we only consider the 134 grid
cells with significant nighttime contribution (at least 24\% with respect to
solar zenith angles greater than 100$^{\circ}$) that were used for the
scientific analysis, this criterion is not always fulfilled. In such cases, we 
iteratively increased the selection radius in steps of 0.1 until the sample 
was large enough. For the example line OH(4-2)P$_1$(1) 
\cite<see also>{noll22}, where the input data set and the resulting intensity
climatology are shown in Figures~\ref{fig:sampling} and \ref{fig:i_2lin}a, the
mean radius was 1.10 for the 134 useful grid points. However, it was only 1.02
for the 113 cells with 100\% nighttime contribution. The mean sample size was
522 with a maximum of 732. Apart from the decrease of the numbers close to
twilight, the sample size shows a remarkable seasonal pattern with maximum
numbers around the equinoxes. These variations reflect changes in the
structure of the X-shooter observing programs, which depend on the seasonal
visibility of the different classes of astronomical objects. As discussed in
section~\ref{sec:data}, the number of data bins is significantly reduced for
OH lines with wavelengths longer than 2.1\,$\mu$m. Nevertheless, the time
resolution and quality of the statistics is only slightly worse. For
OH(9-7)P$_1$(1), we find a mean selection radius of 1.14 and a mean sample
size of 474 for the 134 useful grid points.         

Intensity climatologies relative to the mean as shown in 
Figure~\ref{fig:i_2lin} were corrected for time-specific differences in the 
mean solar radio flux at 10.7\,cm \cite{tapping13}. As OH intensities depend 
on solar activity \cite<e.g.,>{gao16,noll17}, variations in the mean solar 
radio flux by the selection of subsamples can affect intensity climatologies. 
Using the moving 27-day average centered on the day of observation, 
$S_\mathrm{27d}$ (see Figure~\ref{fig:sampling}b), as preferred by
\citeA{noll17}, we found values between 89 and 109 solar flux units (sfu) with
a mean of 99\,sfu for the nighttime grid points related to the representative
example OH(4-2)P$_1$(1). The mean $S_\mathrm{27d}$ values are relatively low in
July/August and relatively high in November/December. In order to minimize the
impact of the solar radio flux on the intensity climatologies, we corrected
the intensities of each grid point to be representative of 100\,sfu, which is
close to the mean value. For this purpose, we performed a linear regression
analysis for the relation between intensity and solar radio flux. The
resulting slopes for the different subsamples were then used for the
correction. The climatologies of these slopes are discussed in
section~\ref{sec:solcycle}. As a standard deviation of 5\,sfu is relatively
small, the correction factors only varied between 0.98 and 1.04 with a
standard deviation of 0.01 for OH(4-2)P$_1$(1). The lowest and highest factors
were found in December and July, respectively. The approach led to a general
decrease of the variance in the selected subsamples. Average reductions
between 1 and 6\% were found for the climatologies of the different lines.

\subsection{Decomposition of Climatologies}\label{sec:decompclim}

For a systematic analysis of the climatologies of different properties for
the whole set of selected OH lines, we used decomposition techniques. First,
we performed the popular principal component analysis (PCA), which is an
orthogonal linear transformation in the feature space that results in a 
new coordinate system with the maximum sample variance along the primary 
axis, the second largest variance along the orthogonal secondary axis, and 
so on. Consequently, a few dimensions are sufficient to describe most of the
variance, which significantly reduces the complexity of the data set. The
most important variability patterns are highlighted, whereas the 
contribution of noise and outliers is minimized. The transformation can be
written as 
\begin{equation}\label{eq:pca}
	\mathbf{T} = \mathbf{X} \mathbf{W}.  
\end{equation}
This matrix equation includes the matrix of the input data $\mathbf{X}$ with 
$n$ rows representing the samples and $p$ columns representing the features.
In our case, $p$ equals 298, i.e. the number of selected OH lines, and $n$ is
134, i.e. the number of useful nighttime data points of the climatologies 
(see section~\ref{sec:calcclim}). Note that the sample mean of each column 
(the individual climatologies) needs to be shifted to zero before the PCA can 
be applied. For a complete transformation, the weight matrix $\mathbf{W}$ and 
the score matrix $\mathbf{T}$ would have sizes of $p \times p$ and 
$n \times p$, respectively. However, as we aim at reducing the 
dimensionality, these matrices are truncated with sizes of $p \times L$ and
$n \times L$, where $L$ is the number of kept dimensions. Our analysis showed
that it is sufficient to choose $L = 2$ as the explained variance is already 
between 89 and 98\% depending on the property (see also 
section~\ref{sec:results}). Moreover, the third and higher components 
indicated that they were strongly affected by variability caused by 
measurement uncertainties. The climatologies of all 298 lines (set to mean 
values of zero) can then be described by the linear combination of two basis 
climatologies provided by the columns of $\mathbf{T}$ and the corresponding 
scaling factors for each line given by $\mathbf{W}$.    

Apart from PCA, we also used nonnegative matrix factorization 
\cite<NMF; e.g.,>{lee99} for the analysis of the different climatologies. NMF
is an approximative dimensionality reduction of the form
\begin{equation}\label{eq:nmf}
	\mathbf{A} \mathbf{B} \approx \mathbf{X},  
\end{equation}
where all matrices have only nonnegative entries. The input data matrix 
$\mathbf{X}$ with $n$ rows and $p$ columns is the same as discussed above.
The result matrices $\mathbf{A}$ and $\mathbf{B}$ have sizes of $n \times L$ 
and $L \times p$, i.e. they provide the basis climatologies and 
line-dependent scaling coefficients for the approximative reconstruction of 
the individual climatologies. Consistent with the PCA, we selected $L = 2$ 
for the simplest description of the variability. Note that the choice of $L$ 
affects the patterns of the basis climatologies, whereas the PCA-related 
components remain fixed independent of $L$. In general, the NMF-related 
results were similar to those of the PCA, even in the case of the solar cycle 
effect, where the possible negative values had to be set to zero for the 
application of the NMF. The corresponding systematic bias was relatively 
small as only a small fraction of slightly negative values contributed to the 
different climatologies. Nevertheless, we preferred the unbiased PCA for this
analysis. The PCA-related results dominate the discussion in
section~\ref{sec:results}. The only exception is section~\ref{sec:intensity}
on intensity climatologies as the NMF-related separation of the two main
contributions was significantly better, which motivated us to also include
that approach in our analysis. The results of both methods are available via
the release of the full data set \cite{noll23ds}.

\subsection{Variability as a Function of Time Scale}\label{sec:vartime}

\begin{figure}
\noindent\includegraphics[width=\textwidth]{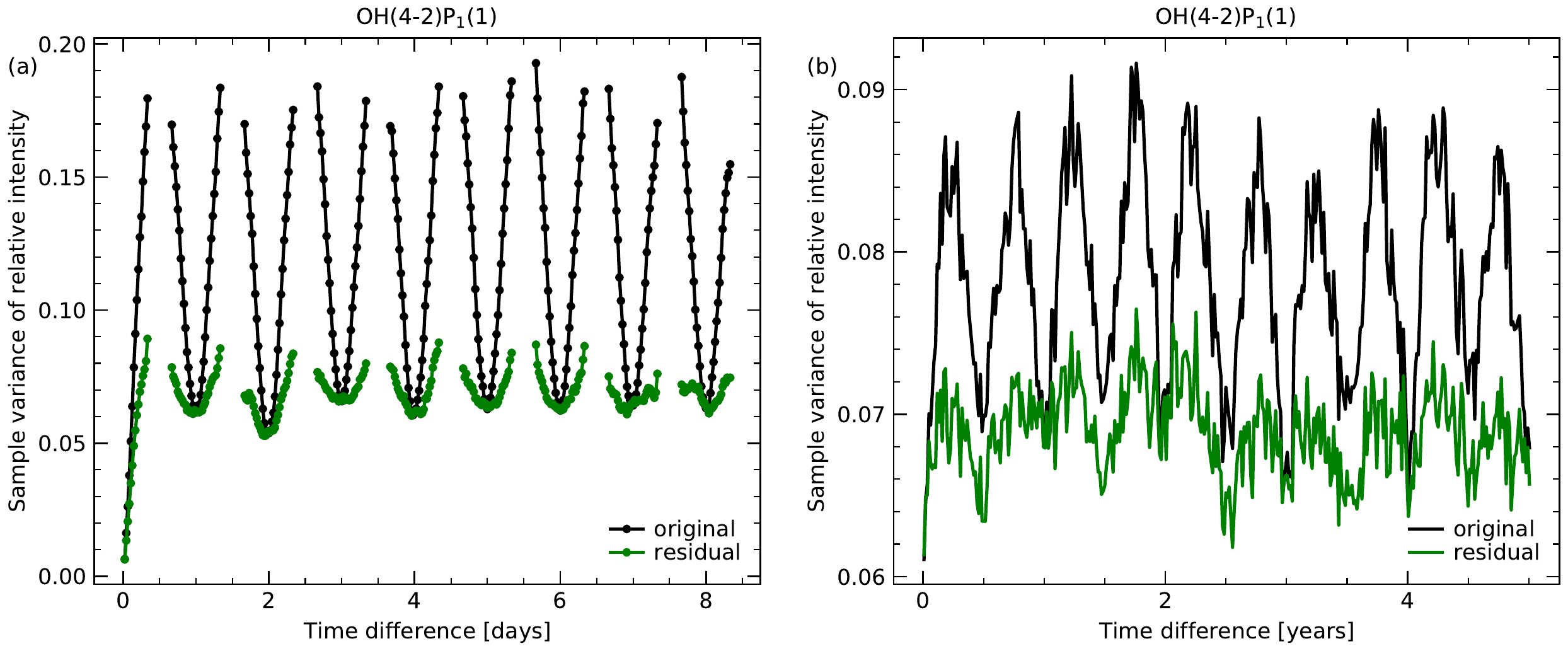}
\caption{Sample variance derived from data pairs as a function of the 
pair-related time difference $\Delta t$ for intensity relative to mean of 
OH(4-2)P$_1$(1) (black) and the latter after subtraction of the 
solar-activity-adapted (section~\ref{sec:calcclim}) climatology in 
Figure~\ref{fig:i_2lin}a (green/gray). (a) shows the variance for $\Delta t$
of the order of hours and days with a time resolution of 30\,min (marked by 
dots), i.e. the step size of the binned input data. Time differences that are
rare due to the restriction to nighttime data are skipped. (b) displays the 
results for $\Delta t$ of multiples of full days up to 5 years. For a 
smoother plot, only the averages of bins with a width of 5 days are shown.}
\label{fig:var_dt}
\end{figure}

Our analysis also comprises a study of the variations which are not covered 
by the 2D climatologies of intensity and solar cycle effect. This residual
variability (section~\ref{sec:residvar}) can be caused by intensity changes 
on different time scales. For the interpretation, it is therefore important 
to understand the contributions of the different wave types depending on the 
climatological grid point. As the X-shooter data set is highly inhomogeneous, 
this analysis requires a robust approach with respect to data gaps. A 
promising statistical method is the study of intensity differences between
all possible combinations of data pairs as a function of the corresponding 
time difference. As the time series consist of bins with a width of half an 
hour (section~\ref{sec:data}), we can assign each pair of relative 
intensities $I_i$ and $I_j$ to a time difference $\Delta t = t_i - t_j$ which
is a whole-number multiple of 30\,min. Then, we can calculate the mean sample 
variance 
\begin{equation}\label{eq:var_dt}
	s^2(\Delta t) = \frac{1}{N(\Delta t)} \sum_{\Delta t, i > j} s_{ij}^2
    \quad \mathrm{with} \quad s_{ij} = \frac{|I_i - I_j|}{\sqrt{2}} 
\end{equation}   
for the $N(\Delta t)$ pairs of each $\Delta t$. Without restrictions in 
$\Delta t$, it would equal the normal sample variance based on deviations 
from the mean value of the full data set. Moreover, the definition has the
advantages that pure noise with a Gaussian distribution causes a constant 
shift of $\sigma^2$ and that a sine wave with the period $T$ results in an 
oscillation between 0 (achieved for integer multiples of $T$) and the squared 
amplitude $a^2$ (shifted by half a period). 

In order to illustrate the approach, Figure~\ref{fig:var_dt} shows 
$s^2(\Delta t)$ for the intensities relative to the mean for our example line
OH(4-2)P$_1$(1) based on the full number of 19,546 useful bins. The function 
is given for relatively short $\Delta t$ up to 8.5 days (a) and longer time
scales up to 5 years with a step size of 5 days (b). For the latter case, 
only $s^2$ for integer multiples of 1 day were averaged. There is a 
dominating oscillation with a period of 24\,h, which reflects the strong
nocturnal trend in Figure~\ref{fig:i_2lin}a. Moreover, a clear semi-annual 
oscillation is visible, which mainly originates from the climatological
pattern at the beginning of the night. As we are primarily interested in the 
residual variability, the 2D intensity climatology for 100\,sfu in 
Figure~\ref{fig:i_2lin}a adapted to the actual $S_\mathrm{27d}$ values 
(section~\ref{sec:calcclim}) was subtracted from the times series. The 
results are also shown in Figure~\ref{fig:var_dt}. As expected, 
$s^2(\Delta t)$ of the corrected time series indicates distinctly lower 
variances. Exceptions in Figure~\ref{fig:var_dt}a are only present for time 
scales of a few hours and multiples of 24\,h. They reveal the importance of 
short-term variations and day-to-day variations of the nocturnal pattern for 
the residual variability. For long time scales, the lowest reduction can be 
seen for multiples of 1 year. Hence, year-to-year variations are probably 
more crucial for the annual oscillation, which dominates the second half of 
the night in Figure~\ref{fig:i_2lin}a.       

The analysis can also be performed for each grid point of the 2D 
climatology. In this way, the contributions of the different time scales to
the residual variability climatology can be studied in detail. As only a 
small fraction of the entire data set is relevant for a specific grid point 
(see section~\ref{sec:calcclim}), there can be a lack of suitable pairs for a 
certain $\Delta t$. This issue affects long time scales in general as well as 
short times scales where the absence of daytime data matters. Good statistics 
are therefore limited to $\Delta t$ of a few hours and those close to low 
multiples of 24\,h. These restrictions still allow the detailed study of the 
impact of short-term variations with time scales shorter than 1 day, which 
constitute the largest contribution to the residual variability as 
Figure~\ref{fig:var_dt}a reveals. For maximum robustness, we measured the
short-term variance as the minimum of $s^2$ for $\Delta t$ of 24 and 48\,h.
In the case of the whole time series of OH(4-2)P$_1$(1), the variance for 
48\,h is clearly lower. This fact suggests that there is a significant 
contribution of the Q2DW, although its activity period is usually only a few 
weeks in summer \cite<e.g.,>{ern13}. Consequently, we can also estimate the 
amplitude of the Q2DW. Assuming a sine wave, we derived $a^2$ by subtracting 
$s^2$ for 48\,h from the mean for 24 and 72\,h. In the case of a negative 
difference, we set $a = 0$. This approach may lose a part of the amplitude as 
the Q2DW period can deviate from 48\,h. \citeA{noll22} found 44\,h for the
covered intervals in 2017 and 2019 at Cerro Paranal. However, the restriction 
to multiples of 24\,h increases the robustness of the statistics and avoids 
possible biases depending on LT due to different influences of daytime 
intervals without observations. Concerning the statistics, $\Delta t = 72$\,h 
is the time difference with the lowest number of data pairs. For the 
nocturnal grid points of the climatology of OH(4-2)P$_1$(1), we find numbers 
between 88 and 300 with a mean of 172, i.e. about one third of the average 
sample size. In conclusion, our analysis allows us to study 2D climatologies 
of short-term variations and Q2DW amplitudes. The corresponding results will 
be discussed in sections~\ref{sec:shortvar} and \ref{sec:2daywave}.

\section{Results}\label{sec:results}

\subsection{Relative Intensity}\label{sec:intensity}

Figure~\ref{fig:i_2lin} shows example climatologies of intensity relative to
the climatological mean for a solar radio flux of 100\,sfu and solar zenith 
angles greater than 100$^{\circ}$. Both example lines belong to the P$_1$ 
branch of the strong OH(4-2) band but show the maximum difference in the upper 
rotational quantum number $N^{\prime}$ of the line set (1 vs. 14). The
reference intensities for the two climatologies are 9.65 and 0.18\,kR,
respectively. Hence, even the high-$N^{\prime}$ line is still relatively
bright. The mean intensity is comparable to those of the green and red atomic
oxygen lines at Cerro Paranal \cite{noll12}. The derived intensity
climatologies appear to be quite robust since the relative root mean square
averaged for the climatologies of the 23 lines with $N^{\prime} \ge 10$ (mean
intensities between 7 and 570\,R) is only 2.5\%, irrespective of possible true
physical differences. The comparison of both climatologies in
Figure~\ref{fig:i_2lin} reveals clear differences at the beginning of the
night, where OH(4-2)P$_1$(1) shows much stronger emission relative to the
mean. The maximum values near both equinoxes (April and October) are hardly
visible in the climatology of OH(4-2)P$_1$(14). The maximum in the second half
of the year even appears to be shifted to November. On the other hand, the
patterns after midnight are more similar. In particular, the maximum in May
before dawn and the minimum in August/September can be found in both
climatologies. A 2D intensity climatology for Cerro Paranal was already shown
by \citeA{noll17} for P-branch lines with low $N^{\prime}$ in OH(6-2). The
summed intensity of these lines indicates a variability pattern that agrees
quite well with our results for OH(4-2)P$_1$(1). The data of \citeA{noll17}
were taken between April 2000 and March 2015 with another VLT spectrograph.
Moreover, similar features as in Figure~\ref{fig:i_2lin}a are also present in
the nocturnal trends and monthly variations of the OH(9-4) band (dominated by
lines with low $N^{\prime}$) measured by \citeA{takahashi98} at Cachoeira
Paulista in Brazil (23$^{\circ}$\,S, 45$^{\circ}$\,W) for the period between
October 1987 and June 1993. The 2D climatology of the OH(8-3) band at Buckland
Park in Australia (35$^{\circ}$\,S, 139$^{\circ}$\,E) for the years 1995 to
2010 from \citeA{reid14} also indicates a rough agreement. Consequently, the
shown intensity climatologies appear to be relatively stable with respect to
the observing period and moderate changes of the latitude. For larger changes
of the latter, there can be significant deviations as a SABER-based study of
the global OH peak emission rates by \citeA{gao10} suggests. The impact of the
latitude is also illustrated by the study of \citeA{takahashi98}, which also
contains results for a site close to the equator (4$^{\circ}$\,S).  

\begin{figure}
\noindent\includegraphics[width=\textwidth]{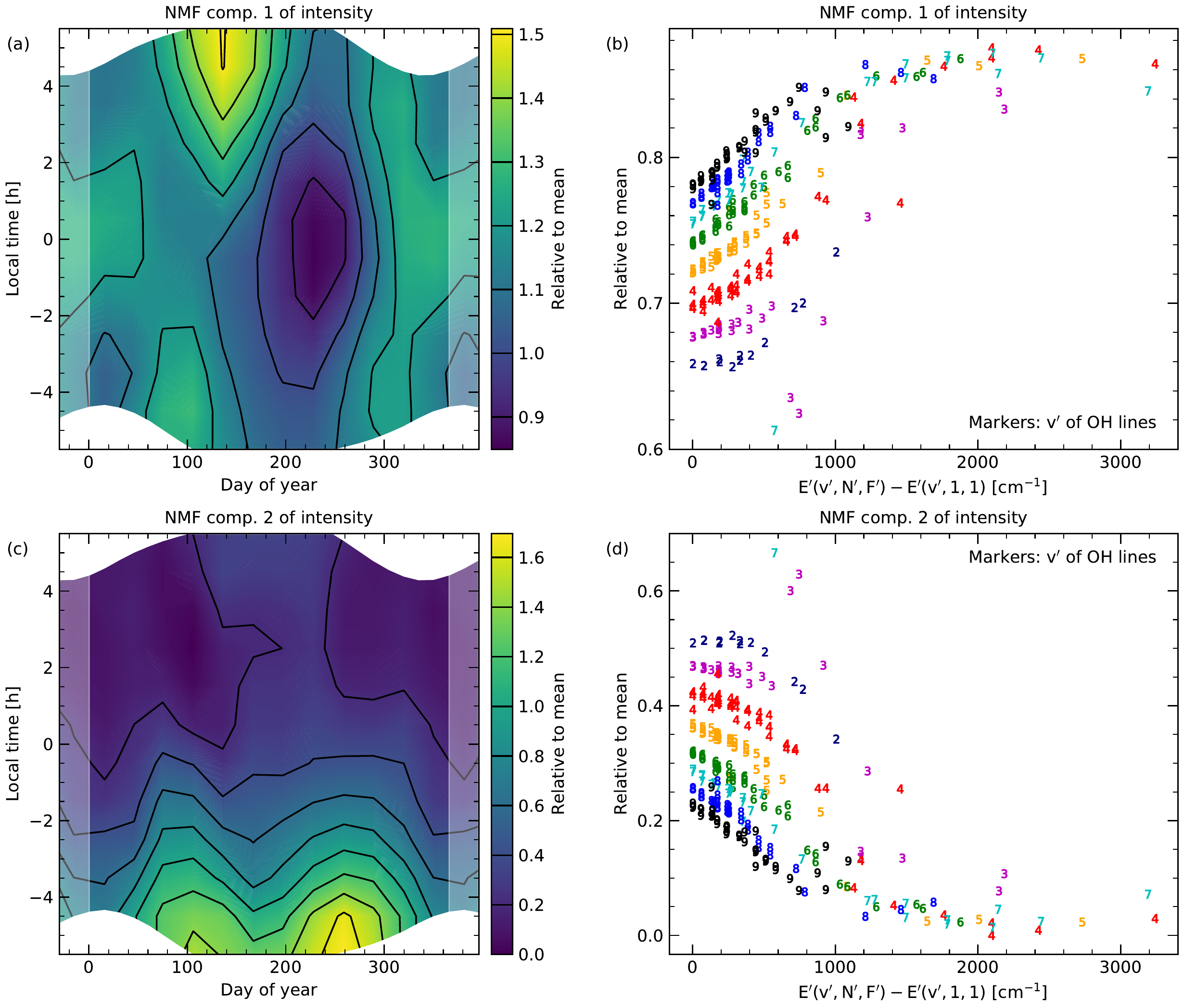}
\caption{Decomposition of climatologies of intensity relative to mean with
nonnegative matrix factorization for two components. The resulting 
climatologies (see also Figure~\ref{fig:i_2lin}) as given by matrix 
$\mathbf{A}$ in section~\ref{sec:decompclim} are shown in (a) and (c) 
and the corresponding coefficients for the 298 considered OH lines from
matrix $\mathbf{B}$ are given in (b) and (d). The number symbols in the 
latter plots indicate the upper vibrational level of the transition 
$v^{\prime}$. The abscissa shows the energy of the upper level of the 
transition minus the lowest energy for the corresponding $v^{\prime}$ in 
inverse centimeters. The additional energy is related to rotation with 
quantum numbers $N^{\prime} > 1$ and/or spin--orbit coupling with quantum 
number $F^{\prime} = 2$ \cite<see>{noll20}.}
\label{fig:i_nmf}
\end{figure}

Our data set particularly benefits from the parallel coverage of hundreds of 
OH lines which show differences in their effective emission heights of 
several kilometers \cite{noll22}. Therefore, more detailed insights into the 
OH-related dynamics will be possible if the climatologies of all 298 lines 
are jointly studied in a systematic way. For this purpose, we applied the 
decomposition techniques that were introduced in 
section~\ref{sec:decompclim}. The first two components of the PCA (matrix 
$\mathbf{T}$ in section~\ref{sec:decompclim}) explain 98.4\% of the full 
variance. The first component is similar to the intensity climatology of 
OH(4-2)P$_1$(1), whereas the second one better agrees with the climatology of
OH(4-2)P$_1$(14). This result confirms that the variabilities of our example 
lines differ relatively strongly with respect to the full line set. The PCA 
was not able to properly separate the high intensities at the beginning of 
the night, which are only visible for one example line, from the other 
variability features that can be found in the data of both lines. As stated 
in section~\ref{sec:decompclim}, we therefore preferred the NMF. It works 
particularly well if a pattern can be reconstructed by summing a few 
nonnegative components. For our analysis, we calculated the decomposition for 
the least complex case, i.e. two components.  

A comparison of the resulting basis climatologies in the left column of
Figure~\ref{fig:i_nmf} (matrix $\mathbf{A}$ in section~\ref{sec:decompclim}) 
with the example cases in Figure~\ref{fig:i_2lin} shows that the NMF clearly 
separates the strongly line-dependent nocturnal trend of decreasing 
intensities from the underlying features that are obviously present in all 
climatologies. The latter are characterized by the first component in 
Figure~\ref{fig:i_nmf}a, which is very similar to the climatology of 
OH(4-2)P$_1$(14). The correlation coefficient $r$ for the climatological grid 
cells with significant nighttime contribution is $+0.94$. For the 
interpretation of this remarkable pattern with the maximum in May at 05:30 LT 
and the minimum in August at 23:30 LT, it is important to know whether it is 
restricted to OH intensities or whether it can also be seen in other 
properties of the mesopause region. The kinetic temperature is a popular 
quantity for the study of mesospheric perturbations. It can be estimated from 
intensity ratios of OH lines if the involved level populations are in local 
thermodynamic equilibrium (LTE), which is best fulfilled for the lowest 
$v^{\prime}$ and $N^{\prime}$ 
\cite{cosby07,kalogerakis18,noll15,noll16,noll20}. We therefore analyzed the 
ratio of the P$_1$ lines of OH(3-1) with $N^{\prime} = 1$ and 2, which are 
frequently used for airglow instruments optimized for temperature 
measurements \cite<e.g.,>{schmidt13}. The climatology of the line ratio 
indicates a good correlation with the first NMF component ($r = +0.83$). 
There are no increased values in the evening as it is typical of individual 
lines with low $v^{\prime}$ and $N^{\prime}$ like OH(4-2)P$_1$(1) 
(Figure~\ref{fig:i_2lin}a). This result is also confirmed by the 2D 
climatology of kinetic temperature based on SABER measurements at 89\,km in 
the region of Cerro Paranal from \citeA{noll19}. The same publication shows 
that similar features are also present for the number density of atomic 
oxygen, which is crucial for the production of OH. Hence, the first NMF 
component is an indicator of general perturbations of the mesopause region. 
As the climatologies amplify variations with fixed time scales of 24, 12, or 
8\,h, solar tides and the seasonal changes of their amplitudes appear to be 
the main source of the variability pattern. The significant impact of tides 
on OH emission was already discussed before 
\cite<e.g.,>{marsh06,takahashi98,zhang01}. It is especially important for low 
latitudes where the migrating diurnal tide that follows the apparent motion 
of the Sun is the most prominent mode. The residual meridional circulation
that can influence the seasonal variability appears to be a minor effect at
low latitudes \cite{marsh06}.        

The second NMF-related basis climatology is shown in Figure~\ref{fig:i_nmf}c.
In each month, the intensity is decreasing in the course of the night with
the highest rates in the evening. Moreover, the pattern indicates a 
semi-annual oscillation with the maximum values near the equinoxes (April
and September). The latter could be related to the ozone number density in
the OH emission layer, which indicates a similar seasonal variation 
\cite{noll19}. As the nocturnal decrease is not visible in the climatologies
of kinetic temperature and atomic oxygen number density shown by
\citeA{noll19}, the phenomenon seems to be linked to the OH-related chemistry.
As already discussed by \citeA{marsh06}, the drop in intensity is probably
caused by the consumption of atomic oxygen, which is mostly produced at
daytime due to photolysis of molecular oxygen. In the denser atmosphere at
lower altitudes, the losses (also by the production of OH via ozone) are
faster. Assuming an exponential decay, \citeA{marsh06} modeled time constants
of 6\,h at 84\,km and 1 day at 88\,km, which indicates a large vertical
gradient of this property. We can also fit exponential functions in the second
NMF component. For the natural logarithm of the values, only a linear
regression analysis needs to be performed. For a better robustness, we only
considered local times until the nocturnal minimum and only grid points with
relative intensities higher than 0.04. The average of all monthly fits amounts
to $3.3 \pm 0.2$\,h, which suggests that the crucial altitudes are probably 1
to 3\,km below 84\,km (if the model of Marsh et al. works for our data). 
Emission at higher altitudes with long decay times appears to mainly 
contribute to the first basis climatology in Figure~\ref{fig:i_nmf}a. The 
second NMF component is therefore strongest for OH lines with the lowest 
effective emission heights, which are about 86\,km on average \cite{noll22} 
but could be about 2\,km lower in the early evening as SABER data for Cerro 
Paranal indicate \cite{noll18a}. Our regression analysis also revealed that 
there might be seasonal variations of the time constant. The values of 
$2.6 \pm 0.1$\,h for austral autumn and $4.0 \pm 0.1$\,h for austral winter 
show the largest discrepancies. 

The scaling factors of the two basis climatologies for all 298 lines are 
provided in the right column of Figure~\ref{fig:i_nmf} (matrix $\mathbf{B}$ 
in section~\ref{sec:decompclim}) as a function of the energy of the upper 
state of the transition reduced by the lowest energy for the corresponding 
$v^{\prime}$, i.e. this energy difference (given in inverse centimeters) 
mainly depends on $N^{\prime}$. Except for a few unreliable outliers, the 
scaling factors indicate a clear transition from climatologies with a slight 
dominance of the second component (decaying daytime population) in the 
evening for the lowest $v^{\prime}$ and $N^{\prime}$ to a contribution of the 
first component (tidal features) of almost 100\% for the highest $N^{\prime}$
irrespective of $v^{\prime}$. For low rotational energies, the different 
$v^{\prime}$ are well separated with larger gaps for lower vibrational 
excitations. In an intermediate zone between 500 to 1300\,cm$^{-1}$, the 
factors of lower $v^{\prime}$ show larger changes, which then leads to the 
vanishing discrepancies for the highest $N^{\prime}$. This distribution is 
very similar to the effective emission heights derived by \citeA{noll22} for 
the same line set as correlation coefficients of $+0.93$ and $-0.92$ for the 
first and second NMF component demonstrate. This excellent agreement supports 
the assumption that the mixing of the two basis climatologies strongly 
depends on the height distribution of the emission. Hence, the differences in 
the coefficients for the studied lines should mostly be caused by the strong 
height dependence of the time constant for the decay of the daytime 
population of atomic oxygen.              

\begin{figure}
\includegraphics[width=20pc]{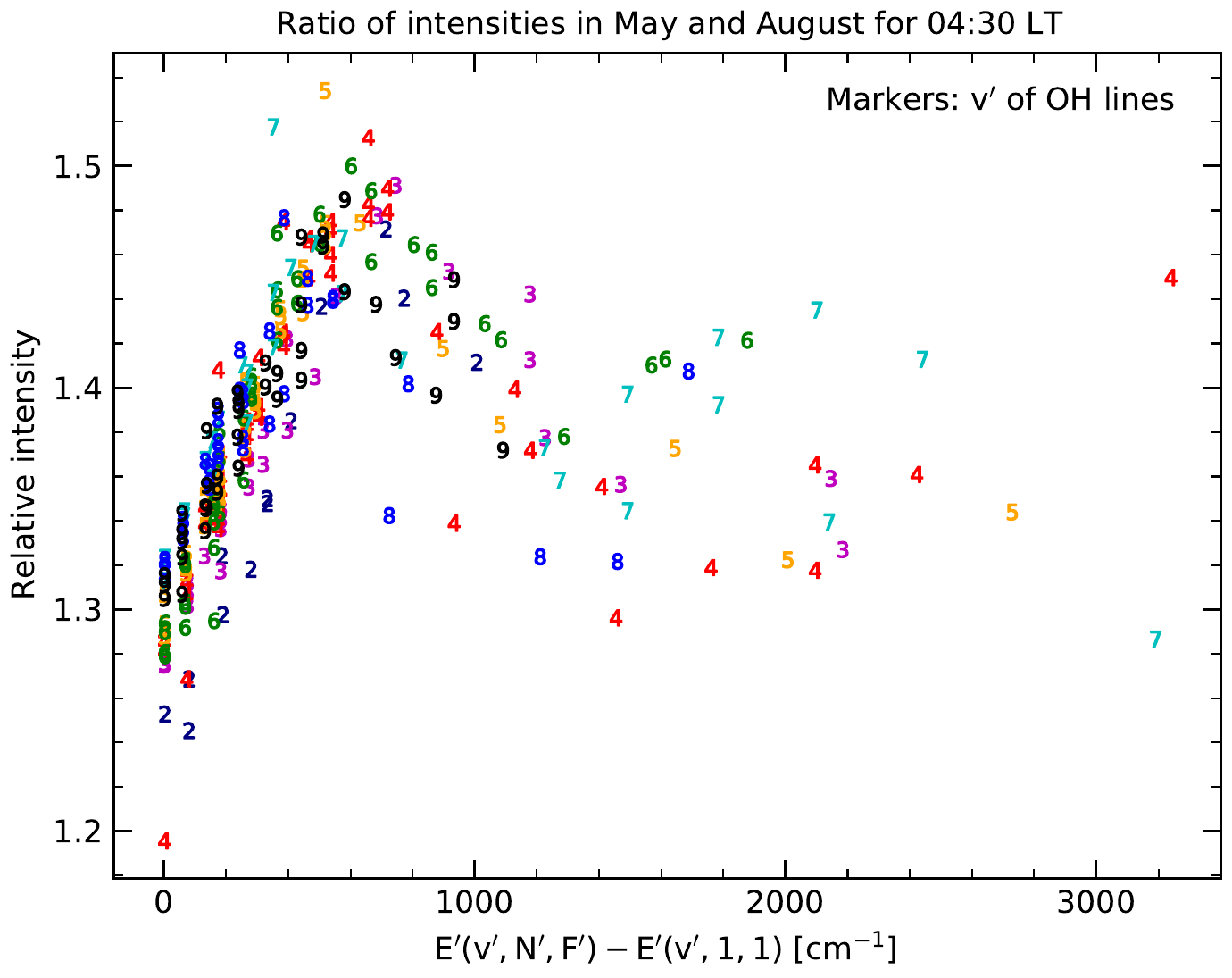}
\caption{Ratio of intensities in May and August for 04:30 LT derived from 
the intensity climatologies of all 298 lines. The upper levels of the 
transitions are characterized by $v^{\prime}$ (markers) and the energy 
additional to the vibrational excitation (abscissa).}
\label{fig:ri_feat}
\end{figure}

While only a relatively narrow altitude range seems to significantly 
contribute to the second NMF component, tidal features should be present at
all altitudes. Hence, the strength of these structures might depend on the 
studied OH line. With the NMF decomposition of the climatologies into two
components, this question cannot be answered. However, we can directly 
measure features in the climatologies for this purpose. As the contribution 
of the decaying atomic oxygen population can be neglected in the morning, 
the best suitable feature is the maximum in May (Figure~\ref{fig:i_nmf}a). 
For an estimate of the strength of this feature, we divided the relative
intensity at 04:30 LT in May by the value for the grid point at the same 
local time in August. This choice is a compromise between high contrast,
good nighttime coverage, and late time. The resulting intensity ratios are 
plotted in Figure~\ref{fig:ri_feat}. The most prominent structure of the
distribution of the 298 data points is the maximum in the energy range from
400 to 800\,cm$^{-1}$. Interestingly, the Q2DWs in 2017 and 2019 investigated 
by \citeA{noll22} showed the largest amplitudes in a similar range. As 
already stated in section~\ref{sec:intro}, this effect can be explained by an 
increased level of variability due to differences in the mixing of cold and 
hot populations caused by variations in the ambient temperature. It seems 
that variations by tides have a similar impact on OH emission. Nevertheless,
the correlation coefficient for the whole set of lines in
Figure~\ref{fig:ri_feat} and the amplitude of the Q2DW event in 2017 is only
$+0.33$. An important discrepancy is the lack of a dependence on $v^{\prime}$
for low $N^{\prime}$. Moreover, the ratios for very high and low $N^{\prime}$
only differ by a few percent. Hence, structures that suggest an impact of the
emission height cannot clearly be identified. The tidal modes that cause the
measured climatological feature appear to affect the different emission
altitudes at a similar time. The effective vertical wavelengths of these
perturbations seem to be relativey long. This interpretation is supported by
the fact that the location of the feature is relatively stable in the two
example climatologies in Figure~\ref{fig:i_2lin}, which correspond to
emissions with an effective altitude difference of about 5.5\,km
\cite{noll22}. If this result was also valid for the other tidal features, the
structures in the first NMF component could be similar for all OH lines.

\subsection{Solar Cycle Effect}\label{sec:solcycle}

\begin{figure}
\noindent\includegraphics[width=\textwidth]{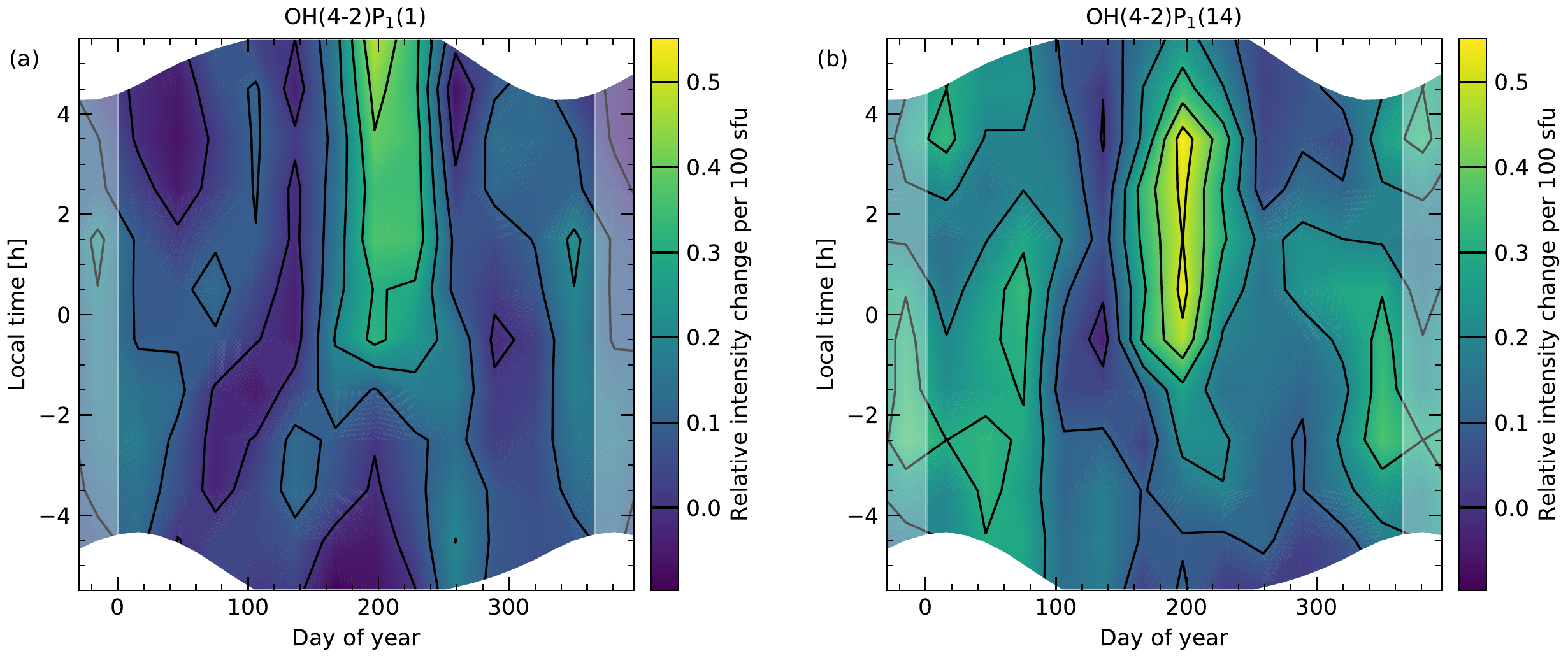}
\caption{Climatologies of solar cycle effect relative to the intensity
climatologies in Figure~\ref{fig:i_2lin} (see caption for plot details) for 
OH(4-2)P$_1$(1) (a) and OH(4-2)P$_1$(14) (b). The climatologies indicate the 
response of OH emission to changes of the solar radio flux averaged for 27
days by 100\,sfu. Each value of the climatological grid is given relative to
the corresponding mean intensity.}
\label{fig:sce_2lin}
\end{figure}

As already described in section~\ref{sec:calcclim}, we analyzed the impact 
of the solar cycle on OH intensities using a linear regression approach with 
the moving 27-day average $S_\mathrm{27d}$ of the solar radio flux at 
10.7\,cm as a proxy. Calculations with an additional linear long-term trend
showed that the latter can be neglected. The ratio of the mean solar cycle
effect (SCE) without and with this trend for all 30\,min bins of all OH lines
did not significantly differ from unity (0.977). Moreover, the long-term trend
was not significant ($+1.2 \pm 0.9$ \% per decade). These results confirm
previous findings \cite{gao16,noll17}. Hence, the subsequent regression
results were obtained with the solar radio flux as the only parameter. Our
data set covers almost the entire solar cycle 24, i.e. the time series starts
and ends with low solar activity as Figure~\ref{fig:sampling}b shows. The
minimum $S_\mathrm{27d}$ for the 30\,min bins in both halves of the interval
are 71\,sfu in October 2009 and 67\,sfu in March 2018. The maximum of 166\,sfu
was achieved in February 2014. Hence, the impact of solar cycle 24 can be
investigated for a range of about 100\,sfu. The mean value also amounts to
about 100\,sfu. The regression analysis was separately performed for the
subsamples of each climatological grid point, which allowed us to study the
influence of the solar activity cycle as a function of LT and month. 

The results for our two example lines relative to the corresponding intensity
climatologies (Figure~\ref{fig:i_2lin}) and for a change of 100\,sfu are 
shown in Figure~\ref{fig:sce_2lin}. The climatologies are remarkable in 
terms of the large inhomogeneities of the SCE. A particularly strong influence
of the solar activity on the OH intensities can be found from midnight to dawn
around July. Then, a relative SCE of more than $+40$\% (with uncertainties
lower than 10\%) per 100\,sfu is possible. For the rest of the climatology,
the response of the OH emission is usually much weaker. Even negative values
cannot be excluded. Only around the turn of the year, the significance of the
regression slopes for part of the grid points is sufficient to safely assume a
positive effect. The major difference between the patterns for both lines is
the location of the maximum with a width of several hours in austral winter.
For OH(4-2)P$_1$(14), it seems to be shifted by a few hours towards earlier
LTs. An exact measurement is difficult as OH(4-2)P$_1$(1) shows the highest
value just before twilight. The latter line also indicates a weaker decrease
of the SCE towards August than in the other case.

Climatologies of the OH-related SCE were rarely investigated before. 
\citeA{pertsev08} compared the months November to January with the period 
from May to July for measurements of the OH(6-2) band in the years from 2000 
to 2006 at Zvenigorod (56$^{\circ}$\,N, 37$^{\circ}$\,E) and found a factor 
of about 2. With an extension of this data set until 2018, \citeA{dalin20} 
obtained a most likely ratio of 1.4 for the periods from 1 October to 31 
March and from 20 May to 15 August. Taking OH(6-3)P$_1$(3) for a comparison, 
we find for the ratio of the intervals from May to August and from October to 
March a value of 1.4, which is close to the results of \citeA{dalin20} if we 
consider a shift in the seasons by 6 months due to the different hemisphere. 
The ratio drops to about 1.2 if April and September are also included in the 
austral winter period. The ratio depends on the selected lines. For our low 
and high $N^{\prime}$ examples in Figure~\ref{fig:sce_2lin}, we obtain 1.7 
and 0.8 for the two intervals with a length of 6 months. \citeA{gao16} used 
global SABER OH data from January 2002 to January 2015 to derive the seasonal 
dependence of the SCE. The result is a maximum in March, which is about 3 
times stronger than the minimum in December. Moreover, July apparently 
belongs to the months with the weakest SCE. These large discrepancies might 
be explained by the comparison of global to local results and the sparse 
LT coverage of the SABER data. For Cerro Paranal, it seems that the local 
times with the largest SCE in austral winter are not well covered 
\cite{noll17}. Finally, \citeA{reid14} investigated the amplitude of the 
solar cycle in OH(8-3) emission at Buckland Park in Australia (see 
section~\ref{sec:intensity}) for different LT intervals. The SCE appears to 
be strongest between 3:00 and 06:00 with a factor of about 1.3 compared to 
the minimum for 0:00 to 03:00. Using OH(8-5)P$_1$(3) and the same LT binning, 
we find the maximum in the latter interval and about twice as strong as the 
minimum in the evening from 18:00 to 21:00. Hence, there is a different 
pattern, which might be related to the differences in the site, the time 
coverage (1995 to 2010), and the approach (harmonic analysis for multiple 
periods).   

\begin{figure}
\noindent\includegraphics[width=\textwidth]{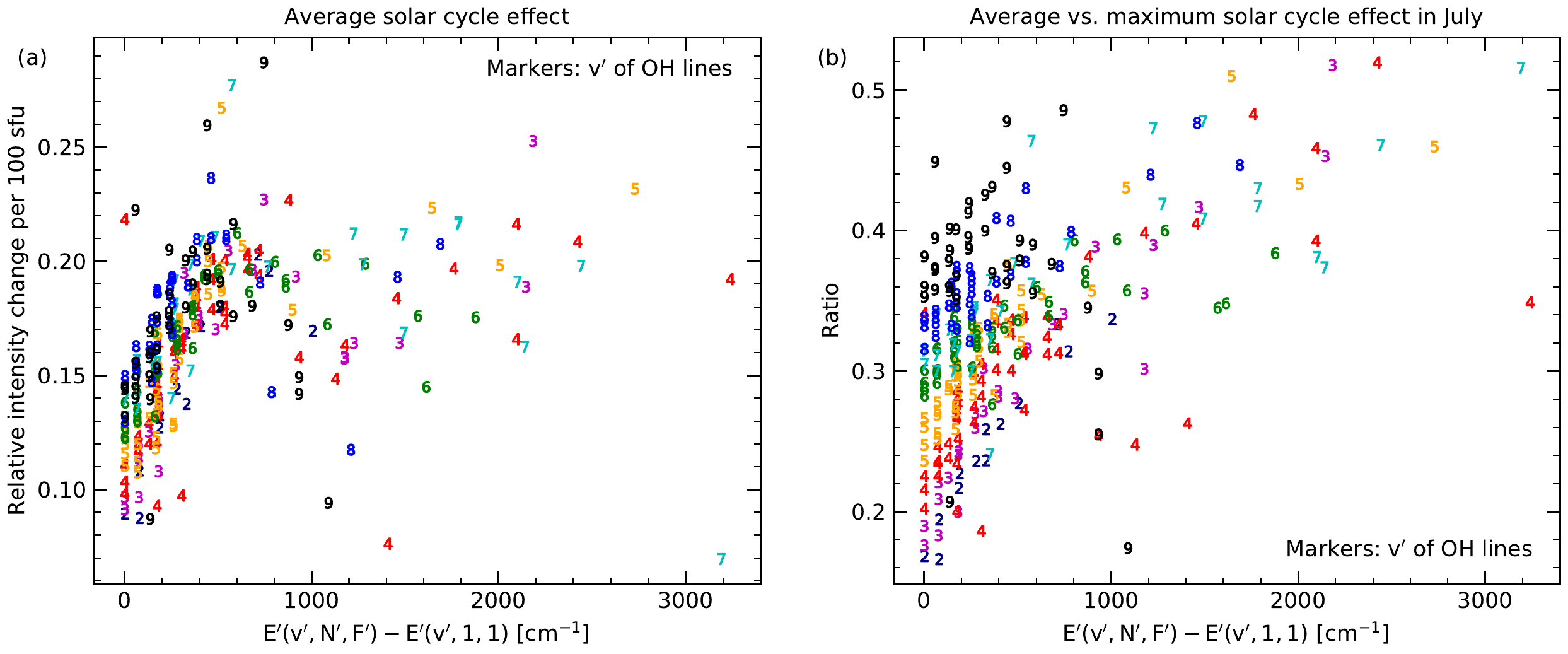}
\caption{Average of solar cycle effect from the corresponding nighttime 
climatology (examples in Figure~\ref{fig:sce_2lin}) for all 298 lines (a). 
The ratio of this average for the entire nocturnal year and the maximum in
July is given in (b) for each line. The plotted line properties are discussed
in the caption of Figure~\ref{fig:i_nmf}.}
\label{fig:sce_eff}
\end{figure}

We also derived the effective SCE for each OH line. For this purpose, we
averaged the relative values for the climatological grid cells under 
consideration of the corresponding nighttime contribution, i.e. the fraction 
of the hour bin with a solar zenith angle greater than 100$^{\circ}$. This 
approach minimizes the impact of inhomogeneities in the time series and the
structure of the intensity climatologies. The results are shown in 
Figure~\ref{fig:sce_eff}a. Although there is some scatter by measurement
uncertainties, which obviously have a larger impact on the related regression
analysis than on the simple averaging of intensities, a clear trend is visible
for rotational energies up to about 600\,cm$^{-1}$. There, the average SCE
increases with higher $v^{\prime}$ (at least up to 8) and $N^{\prime}$. For
higher energies, the effect seems to be relatively constant. Even a slight
decrease is possible. The reliable data points are probably located in the
range from 8 to 23\% per 100\,sfu. The strong dependence of the effective SCE
on the OH line parameters was not seen before. \citeA{pertsev08} stated that
the SCEs for OH bands with $v^{\prime}$ between 3 and 9 were similar at
Zvenigorod. Solar forcing for OH emission at Cerro Paranal was already
investigated by \citeA{noll17} based on optical spectra for the period from
April 2000 to March 2015. The results were restricted to summed intensities of
lines with low $N^{\prime}$ of bands with $v^{\prime}$ between 5 and 9. All bands
showed relative SCEs of about 16\% per 100\,sfu with uncertainties much larger
than the discrepancies. A mild trend could be possible for the two OH channels
of SABER \cite{russell99}. Global data suggest a ratio of about 1.1 for the
relative SCEs of the channels centered on 2.1 and 1.6\,$\mu$m \cite{gao16}
with effective $v^{\prime}$ of 8.3 and 4.6, respectively \cite{noll16}.
\citeA{noll17} also presented SABER results for the region around Cerro
Paranal. For the years from 2002 to 2015, the corresponding ratio derived from
$14.5 \pm 1.3$ and $12.1 \pm 1.5$ \% per 100\,sfu is 1.2 but with relatively
high uncertainties. The obvious main reason for the lack of strong differences
is the fact that only lines with low $N^{\prime}$ significantly contributed and 
the range of $v^{\prime}$ was limited. On the other hand, 
Figure~\ref{fig:sce_eff}a suggests that SCEs that were previously estimated 
are in good agreement with our results (despite the differences in the 
samples) if we focus on the relevant lines. 

For a better understanding of the line dependence of the SCE, we also derived
the maximum amplitude for each line in July, i.e. the month with the 
strongest positive response. The results (not plotted but available via the
data release) are very different from those of the effective SCE. With a 
range of the reliable values from 37 to 67\% per 100\,sfu, the 
maximum-to-minimum ratio is distinctly smaller than for the effective SCE. As
the highest values are present in the energy range from 300 to 
800\,cm$^{-1}$, there is a clear similarity to the distribution for the 
intensity ratio plotted in Figure~\ref{fig:ri_feat} and the amplitude of the 
Q2DWs studied by \citeA{noll22}. Consequently, the maximum SCE also appears 
to be significantly affected by the mixing of cold and hot populations. 
Interestingly, the amplitude increases with decreasing $v^{\prime}$ for low 
$N^{\prime}$, which is contrary to the effective SCE in 
Figure~\ref{fig:sce_eff}a but agrees with the results for the Q2DW in 2017. 
In the latter case, the corresponding correlation coefficient of $+0.74$ is 
therefore relatively high. The discrepancies between the effective and the 
maximum SCE are visualized in Figure~\ref{fig:sce_eff}b, which shows the 
ratio of both quantities for all OH lines. Starting with a minimum of about 
0.17 for $v^{\prime} = 2$ and $\Delta E^{\prime}$ near 0\,cm$^{-1}$, the 
ratio strongly increases for higher $v^{\prime}$ (up to 0.38 at 0\,cm$^{-1}$) 
and higher $N^{\prime}$ (average of 0.43 for $\Delta
E^{\prime} > 1500$\,cm$^{-1}$). The bump related to the population mixing has
obviously vanished, i.e. the effective and maximum SCE appear to be affected 
by this feature in a similar way. The remaining probably monotonic increase 
with a flattening for the highest rotational energies is reminiscent of the 
distribution of the effective emission height of the investigated lines 
\cite{noll22}. The correlation coefficient for the comparison of
the mean-to-maximum ratio with these heights is $+0.80$ despite several
outliers in Figure~\ref{fig:sce_eff}b. Hence, the ratio is mainly a function
of the emission height. At least in part, this might be explained by the
location of the maximum SCE with respect to LT as shown in
Figure~\ref{fig:sce_2lin}. While the range of increased SCE in July is well
covered by the nighttime climatology of OH(4-2)P$_1$(14), the largest values
for OH(4-2)P$_1$(1) are achieved before dawn. In the latter case, the
favorable conditions for a strong effect may extend to later LTs. Hence, the
restriction to nighttime observations seems to cause an increasing loss of
times with strong solar forcing for OH lines with lower effective emission
heights, which then contributes to the observed wide spread of the effective
SCEs.  

\begin{figure}
\noindent\includegraphics[width=\textwidth]{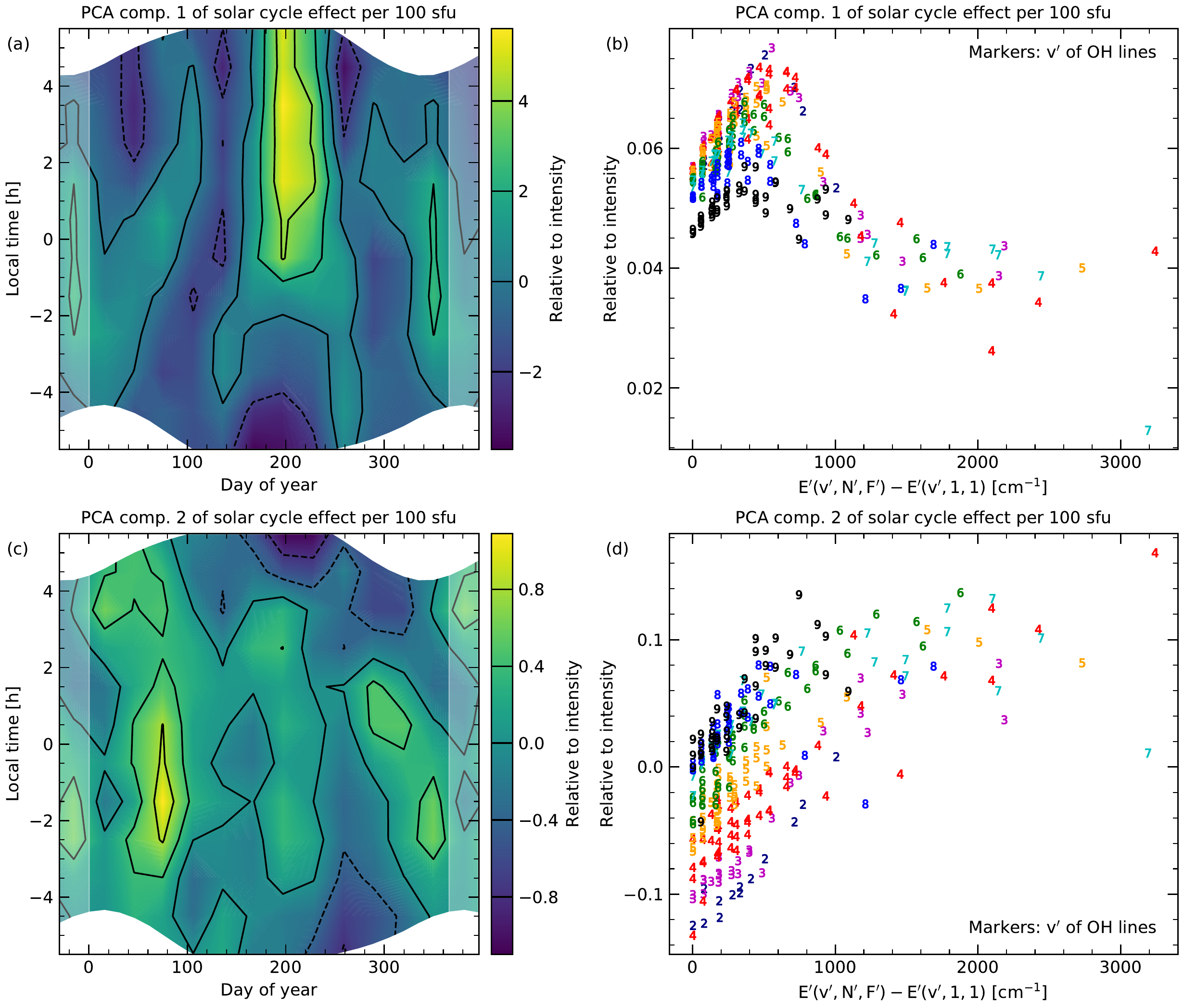}
\caption{Decomposition of climatologies of solar cycle effect (examples in 
Figure~\ref{fig:sce_2lin}) with principal component analysis for two 
components. The plots in the left and right column correspond to the content
of matrices $\mathbf{T}$ and $\mathbf{W}$ described in 
section~\ref{sec:decompclim}. The plot details are similar to those of 
Figure~\ref{fig:i_nmf}.}
\label{fig:sce_pca}
\end{figure}

An alternative approach for the analysis of the climatologies is the use of
decomposition techniques as described in Section~\ref{sec:decompclim}. As 
the SCE can also be negative, we prefer the PCA here. The results for the
first two components, which explain 90.3 and 3.6\% of the variance, are shown
in Figure~\ref{fig:sce_pca}. The first basis climatology in (a) (first column
of matrix $\mathbf{T}$ in section~\ref{sec:decompclim}) is an intermediate 
case compared to the patterns for the two example lines displayed in 
Figure~\ref{fig:sce_2lin}. The corresponding scaling factors for
the individual lines in (b) (first column of $\mathbf{W}$) show a clear bump 
at intermediate rotational energies. Hence, this is another confimation of 
the increased impact of temperature variations for OH rotational levels with 
similar contributions of cold and hot populations. The entire distribution of 
the data points is very similar to the one for the Q2DW in 2017 shown by 
\citeA{noll22}. Benefiting from the noise-reducing properties of the PCA, 
the correlation coefficient is $+0.96$. It is remarkable that two phenomena 
which are related to time scales that differ by several orders of magnitude 
can produce such a similar response of the OH emission intensity. The second 
PCA component in (c) (second column of $\mathbf{T}$) tends to show positive 
values in the middle of the night and exhibits the most negative values close 
to the morning twilight in July and August. Thus, positive scaling factors 
lead to a shift of the feature with the strongest SCE towards earlier LTs in 
the combined climatologies. The coefficients in (d) (second column of 
$\mathbf{W}$) therefore represent a measure of the shift in LT for each OH 
line. The data distribution supports our conclusion that the high SCEs in 
austral winter are present at later times for decreasing $v^{\prime}$ and 
$N^{\prime}$. Moreover, this distribution is highly correlated with the 
effective emission heights ($r = +0.89$), which demonstrates that the shift 
of the SCE features is primarily altitude dependent. As the representative 
LTs are earlier for higher altitudes, the SCE appears to be influenced by 
upward-propagating perturbations. As we investigate climatologies, the 
formation of a robust pattern is only imaginable in the case of tides. Hence, 
the impact of tides on the mesopause region appears to affect the sensitivity
of OH emission to solar forcing. As the shifts could amount to several hours, 
the vertical wavelengths of the relevant tidal modes need to be relatively 
short. We will provide a more quantitative discussion of this topic in 
section~\ref{sec:2daywave}.  

Changes in the solar activity affect the daytime production of atomic oxygen 
and hydrogen by photolysis as well as the energy input into the atmosphere
indicated by the ambient temperature \cite<e.g.,>{beig08,marsh07}. Both 
effects modify the OH intensity. The impact of the solar cycle on the 
temperature at OH emission altitudes can be tested separately via the ratio 
of the OH(3-1)P$_1$(1) and OH(3-1)P$_1$(2) intensities (see 
section~\ref{sec:intensity}). The corresponding SCE climatology shows a very 
similar seasonal pattern as in the case of the intensity of individual lines.
Although the lines have low $v^{\prime}$ and $N^{\prime}$, the maximum effect 
occurs shortly after midnight in austral winter, which better matches lines 
like OH(4-2)P$_1$(14) (Figure~\ref{fig:sce_2lin}). This result supports our 
explanation of the shifts of the SCE pattern in LT direction since the 
effective height for rotational temperature changes tends to be several 
kilometers higher than the effective height for intensity variations 
\cite<e.g.,>{noll22,swenson98}. The temperature changes are weighted by the 
OH intensity profile, whereas the intensity changes usually maximize in the 
lower part of the layer due to the steepening of the atomic oxygen gradient 
\cite<e.g.,>{smith10}. With respect to the presence of atomic oxygen at
relatively low altitudes, the solar activity cycle seems to have a similar
impact as waves with much shorter time scales.

\subsection{Residual Variability}\label{sec:residvar}

\begin{figure}
\noindent\includegraphics[width=\textwidth]{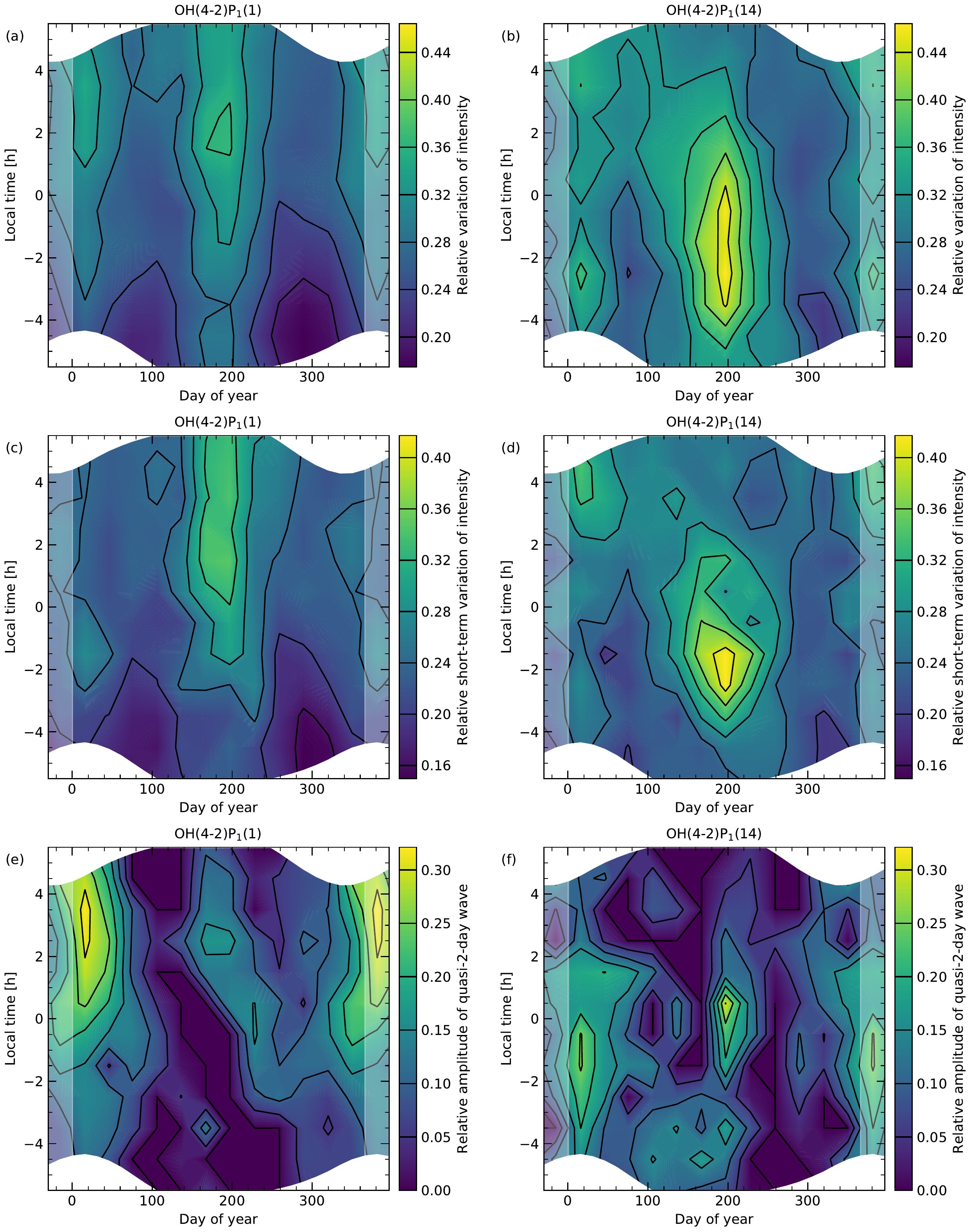}
\caption{Climatologies of residual variability (standard deviation; top row), 
short-term variability (residual variations for time scales less than a day;
middle row), and amplitude of quasi-2-day wave (bottom row) relative to the 
intensity climatologies in Figure~\ref{fig:i_2lin} (see caption for plot 
details) for OH(4-2)P$_1$(1) (left column) and OH(4-2)P$_1$(14) (right 
column).}
\label{fig:di_2lin}
\end{figure}

We also investigated the variability of OH line intensities which cannot be
explained by the average climatologies discussed in 
section~\ref{sec:intensity} and those related to the SCE presented in 
section~\ref{sec:solcycle}. For this purpose, we calculated a line-specific
model intensity for each 30\,min bin depending on the LT interval, month, and
the deviation of the solar radio flux $S_\mathrm{27d}$ from 100\,sfu using 
the corresponding climatologies for relative intensity and SCE. The model 
values were then subtracted from the measured relative intensities. 
Climatologies of the residual variability can now be derived by the 
calculation of the standard deviation of the corrected relative intensities 
for the subsamples related to each climatological grid point (see 
section~\ref{sec:calcclim}). For our two example lines, the results are
provided in the top row of Figure~\ref{fig:di_2lin}. Both climatologies 
show the highest values in June and July and a second maximum around 
January. The main discrepancy is the LT range with the highest residual 
variability relative to the climatological mean. The later maximum for 
OH(4-2)P$_1$(1) qualitatively agrees with the results for the SCE
(Figure~\ref{fig:sce_2lin}). 
  
As already discussed in section~\ref{sec:vartime}, we analyzed the time
dependence of the residual variability by means of the calculation of the
sample variance for data pairs with the same time difference. With this 
approach, it was possible to derive climatologies for short-term variability 
with time scales shorter than a day and the amplitude of the Q2DW for each OH
line. As we only needed to use the frequently occurring time differences of 
24, 48, and 72\,h for the calculation of these properties, the results are 
relatively robust. We discuss them in the following.

\subsubsection{Short-Term Variability}\label{sec:shortvar}

As demonstrated by Figure~\ref{fig:var_dt} for OH(4-2)P$_1$(1), the variance 
in the corrected intensities is dominated by short time scales up to a few 
hours. Using the minimum of the variance values for 24 and 48\,h as a 
measure, the short-term variability corresponds to 75\% of the total residual
variance in the case of OH(4-2)P$_1$(1) and 70\% in the case of 
OH(4-2)P$_1$(14). These percentages are the averages for the nighttime 
climatological grid. The individual grid points for OH(4-2)P$_1$(1) reveal a
clear seasonal variation with the highest and lowest fractions in austral 
winter (84\% in August) and summer (61\% in January), respectively. The 
absolute minimum of 47\% at around 03:30 LT in January matches the summer 
maximum of the residual variability. As a consequence, the resulting 
short-term variability in Figure~\ref{fig:di_2lin}c only indicates one 
pronounced maximum in the second half of the night in June and July. The 
secondary maximum in January has mostly vanished. For the weak 
OH(4-2)P$_1$(14) line, the variance fractions are relatively noisy. 
Nevertheless, relatively low variance ratios in the evening in summer lead to
the weakening of the summer peak in Figure~\ref{fig:di_2lin}d. Again, there 
is only one dominating climatological maximum in winter but earlier in the 
night compared to OH(4-2)P$_1$(1). The exact shape of this structure remains 
unclear due to the relatively high uncertainties with respect to the variance 
fraction. 

Our results for the seasonal pattern can be compared to a study of the 
variance of the intensity of the OH(6-2) band measured at El Leoncito in
the Argentinian Andes (32$^{\circ}$\,S, 69$^{\circ}$\,W) by \citeA{reisin04}.
After the removal of diurnal and semidiurnal tidal modes by means of a 
fitting procedure, the authors also obtained a clear primary maximum in June 
and July, and a weaker secondary one in December and January. This pattern,
which was explained by variations in the gravity wave (GW) activity, is also 
present in the corresponding results for OH(6-2)-based rotational 
temperatures, which is consistent with our climatology for the ratio of 
OH(3-1)P$_1$(1) and OH(3-1)P$_1$(2) (see section~\ref{sec:intensity}) that 
indicates a similar LT dependence of the winter maximum as in the case of the
short-term intensity variance for the individual lines. In addition, 
significantly increased GW activity in austral winter was found by 
\citeA{alexander15} in SABER temperature fluctuations with vertical 
wavelengths between 5 and 20\,km for the region around the Andes between 
29$^{\circ}$ and 36$^{\circ}$\,S in the mesosphere and stratosphere, i.e. at 
significantly lower altitudes than those related to the OH emission. 
SABER-based global maps of gravity wave amplitude and momentum flux at 30\,km
\cite{ern18,preusse09} also reveal a winter maximum at Cerro Paranal. The 
increased activity seems to be connected to the GW hot spot in the southern 
Andes, i.e. the winter polar vortex and orographic forcing obviously play a 
role. According to the maps, the shallow summer maximum is probably related 
to GWs forced by deep convection, which have a hot spot east of the Andes at 
low southern latitudes in summer. At least for short-period waves with 
periods of 5 to 10\,min, this interpretation seems to be supported by 
broad-band OH airglow imaging at Cerro Pach\'on in Chile (30$^{\circ}$\,S, 
71$^{\circ}$\,W) studied by \citeA{cao22}. The observations show preferential
propagation directions towards the south (and also west) in austral summer 
and towards the north (and also east) in winter. Filtering effects
\cite<e.g.,>{fritts03} certainly contributed to this pattern as the wind
measured by a meteor radar at the same site tended to indicate opposite
directions, i.e. critical similarities in speed and direction were reduced.
According to \citeA{cao22}, the majority of the detected waves was probably
not directly propagating from the tropospheric hot spots. Either wave ducting
in the mesopause region \cite<e.g.,>{walterscheid99} or secondary wave
generation related to wave breaking at lower levels should be crucial. The
importance of secondary waves is also discussed by other studies related to 
the same geographical region \cite{liu19,vadas19,vargas16}. As indicated by 
Figure~\ref{fig:var_dt}, most of the measured variance is related to
perturbations with periods of hours, i.e. significantly longer than for
the waves studied by \citeA{cao22}. Long-period GWs tend to propagate mainly 
horizontally and can therefore be detected far from the source region
for favorable atmospheric conditions \cite<e.g.,>{fritts03}. Hence, it is
more likely to observe waves at Cerro Paranal that directly originate from 
the tropospheric source regions than in the case of short-period waves. Note
that such discrepancies in the propagation properties can contribute to 
period-dependent differences in the seasonal wave activity 
\cite<e.g.,>{sedlak20}.  

In conclusion, the relatively strong short-term variations in the X-shooter
OH data in June and July appear to be mainly produced by primary or 
secondary GWs related to the winter hot spot south of Cerro Paranal. 
Nevertheless, the details of the generation, propagation, and filtering of 
the relevant waves remain relatively uncertain. The OH intensity variance for
time scales of hours may be affected by day-to-day changes in the tidal 
pattern. However, such changes can be forced by varying interactions with GWs
\cite<e.g.,>{fritts03,smith12}, i.e. the origin of the observed OH intensity 
variance would still be gravity waves. The scenario of strong interactions 
between tides and GWs seems to be supported by the fact that the enhanced 
climatological short-term variations in austral winter in 
Figure~\ref{fig:di_2lin} are obviously embedded in the area of the intensity 
climatologies where the strongest tidal variations as indicated by 
Figure~\ref{fig:i_nmf}a are present. Moreover, there is an interesting 
similarity in the climatologies of the short-term variations and the solar 
cycle effect, which also shows the largest effects in the second half of the 
night in winter in the case of OH(4-2)P$_1$(1) (Figure~\ref{fig:sce_2lin}a). 
Although the maximum SCE appears to be later in LT (a few hours) and day of 
year (a few weeks), the correlation coefficient for these well-defined 
features of similar shape is still 0.53. Hence, it could be that the 
increased SCE is related to enhanced GW activity. The stronger vertical 
transport by the wave-induced perturbations might increase the sensitivity of
the OH production and emission to the atmospheric changes related to solar 
activity (such as the increased production of atomic oxygen).

\begin{figure}
\noindent\includegraphics[width=\textwidth]{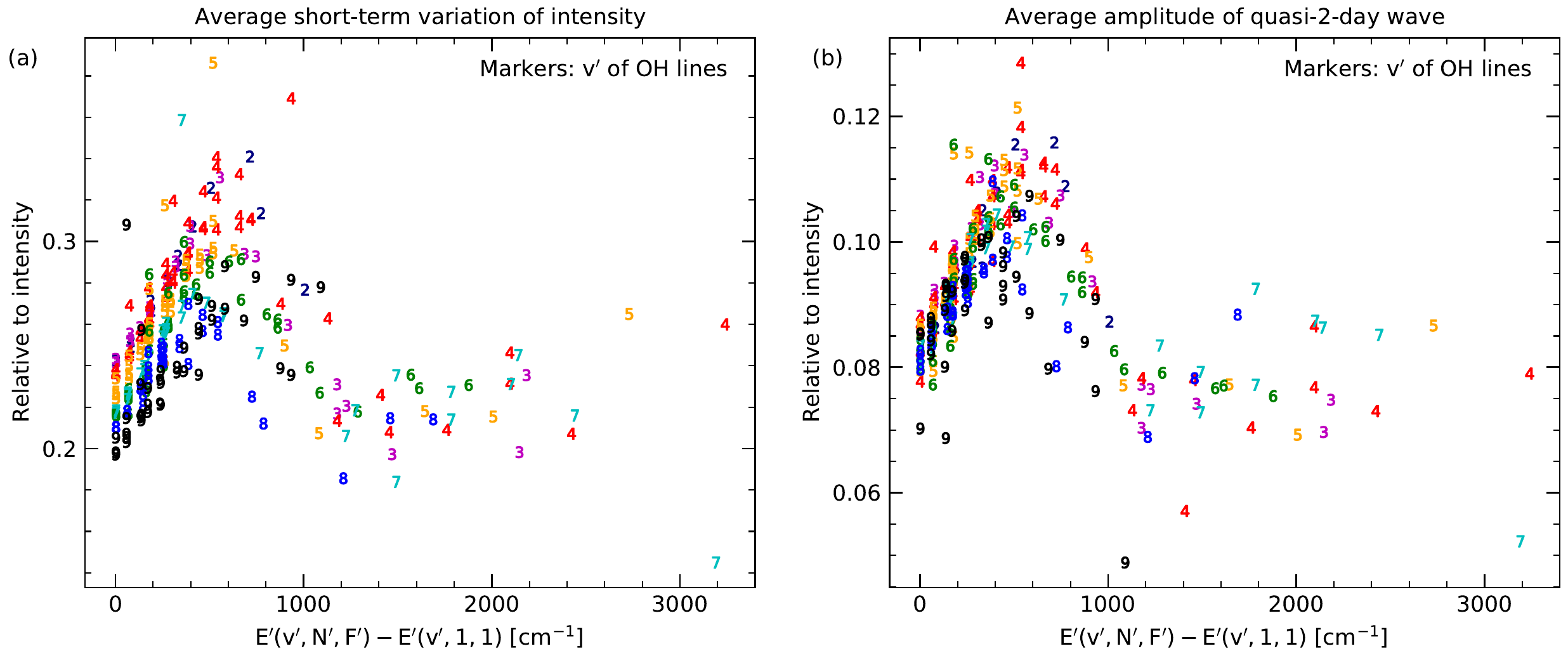}
\caption{Average of short-term variability (a) and Q2DW amplitude (b) from 
the corresponding nighttime climatology (examples in 
Figure~\ref{fig:di_2lin}) for all 298 lines. The plotted line properties are 
discussed in the caption of Figure~\ref{fig:i_nmf}.}
\label{fig:di_eff}
\end{figure}

In the following, we discuss the dependence of the short-term variations on
the line parameters. Figure~\ref{fig:di_eff}a shows the effective relative
standard deviations for the entire nighttime climatologies. The plot was
produced in the same way as for the SCE in Figure~\ref{fig:sce_eff}a. The
distribution shows the typical features related to the mixing of cold and
hot rotational populations. Hence, there is a good positive correlation of
$r = +0.77$ with the amplitude of the Q2DW from 2017 \cite{noll22}. On the
other hand, $r$ is only $+0.28$ in the case of a comparison with the 
SCE-related Figure~\ref{fig:sce_eff}a. The discrepancies seem to be related
to the location of the winter maximum in the climatologies. As discussed in
section~\ref{sec:solcycle}, there appears to be a significant loss of high 
SCE values for OH levels with low $v^{\prime}$ and $N^{\prime}$ due to the
morning twilight. This issue is less critical for the short-term variations,
which show their highest values closer to midnight (e.g. 
Figures~\ref{fig:sce_2lin}a and \ref{fig:di_2lin}c). In agreement with this
assumption, a high $r$ of $+0.83$ is found if the short-term variations are 
only compared with the maximum SCE in July.

\begin{figure}
\noindent\includegraphics[width=\textwidth]{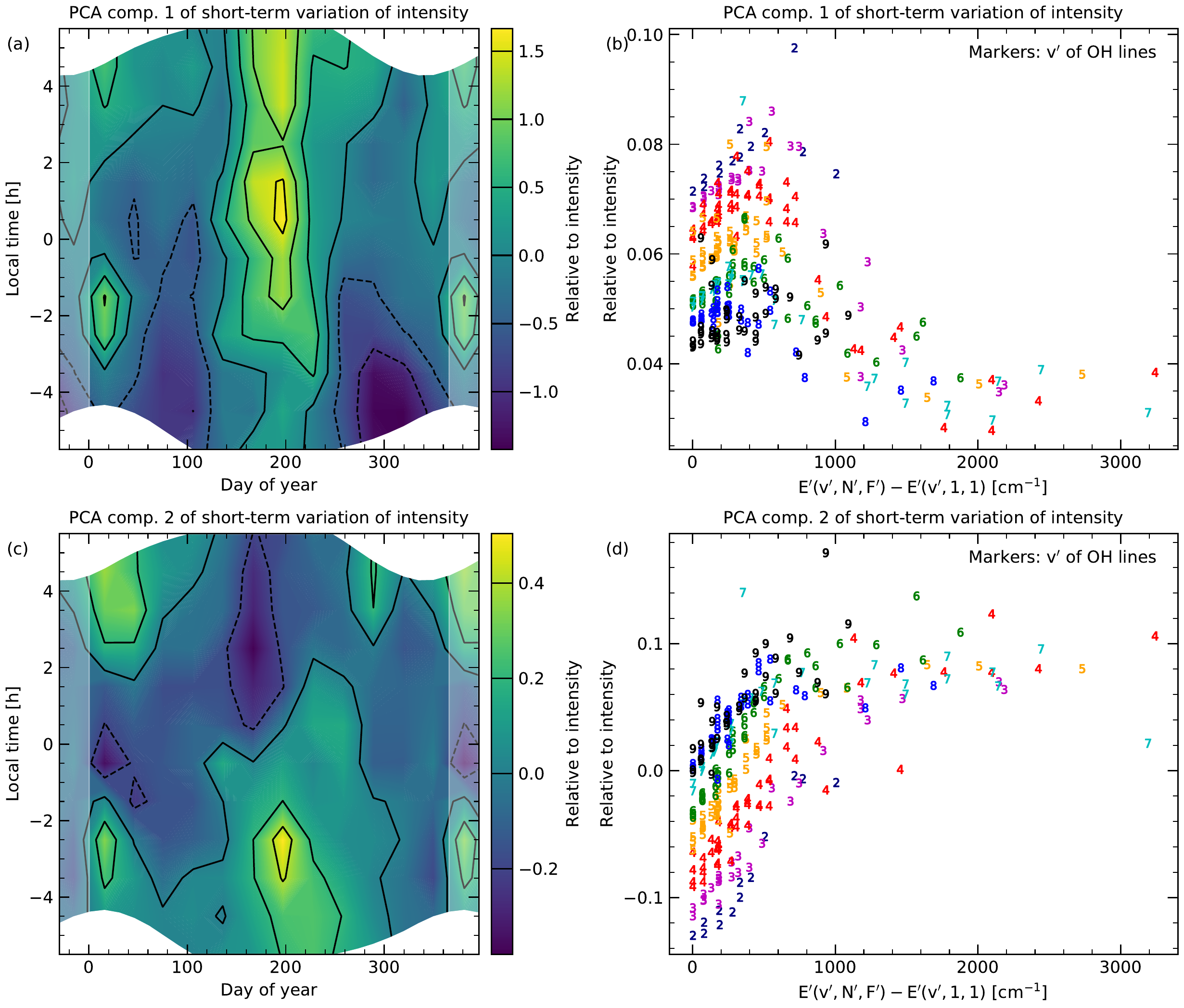}
\caption{Decomposition of climatologies of short-term variability (examples 
in Figure~\ref{fig:di_2lin}) with principal component analysis for two 
components. The figure is similar to Figure~\ref{fig:sce_pca}.}
\label{fig:disw_pca}
\end{figure}

In Figure~\ref{fig:disw_pca}, we show the PCA results for the climatologies
of the short-term variations. The first and the second component, which 
explain 83.6 and 5.3\% of the variance, show similar structures as those
for the SCE. The scaling factors of both components in the right panels of
Figure~\ref{fig:disw_pca} correlate with the corresponding ones for the SCE in
Figure~\ref{fig:sce_pca} with $r$ of $+0.81$ and $+0.94$, respectively. The
first value seems to be lowered by the relatively large 
spread with respect to the vibrational levels in Figure~\ref{fig:disw_pca}b. 
This also causes a relatively strong negative correlation with the effective 
emission heights of $-0.80$ (compared to $+0.85$ for the amplitude of the 
Q2DW in 2017). However, the second component still shows a distinctly 
stronger dependence on the emission height ($r = +0.89$) in agreement with 
the results for the SCE. Consequently, the PCA reveals that the line 
dependence of the short-term variations can also mainly be explained by a 
combination of an amplitude increase for intermediate $N^{\prime}$ due to an 
increased sensitivity to temperature variations (population mixing) and a 
height-dependent shift of the climatological pattern in LT direction. The 
latter is related to a basis climatology for the second component 
(Figure~\ref{fig:disw_pca}c) with an earlier decrease of the values in 
winter (in the middle of the night) than in the case of the SCE (around 
04:00 LT, Figure~\ref{fig:sce_pca}c). If we explain the short-term 
variations as mainly caused by GWs as discussed before, these results imply 
that the presence of GW features in OH emission depends on LT and height. As
the maximum range of effective emission heights for our line set is about
8\,km on average \cite{noll22} with similar layer widths 
\cite<e.g.,>{baker88,noll16}, possible changes in the GW sources and 
filtering effects below the mesopause region do not appear to be favorable 
explanations. Variations in the location of critical layers important for 
wave breaking and wave ducts may play a role. However, as already discussed 
in section~\ref{sec:intensity} with respect to the intensity climatologies, 
especially the vertical distribution of atomic oxygen can cause strong 
line-dependent differences. In this context, height-dependent variations of 
the concentration (accompanied by temperature changes) by independent 
perturbations seem to be important. Considering the amplification of waves 
with periods of integer fractions of a day in the climatologies and the shift
of the pattern towards earlier LTs (up to several hours) with increasing 
height, rising solar tides could be crucial again. They appear to be able to 
alter the OH-related chemistry in a way that significantly affects the 
sensitivity of OH-based observations of GW activity.

\subsubsection{Quasi-2-Day Wave}\label{sec:2daywave}

As described in section~\ref{sec:vartime}, we can also derive 2D 
climatologies of the relative amplitude of the Q2DW based on the differences
of the residual variances for time differences of 2 and 1 plus 3 days. Hence,
we can extend the investigation of Q2DW activity in January/February 2017
(eight nights) and January 2019 (seven nights) \cite{noll22} by a more
statistical approach for the entire data set. For our two example lines,
the results are shown in the bottom row of Figure~\ref{fig:di_2lin}. The
climatologies display clear maximum values in January at about 03:00 LT
for OH(4-2)P$_1$(1) and about 23:00 LT for OH(4-2)P$_1$(14), i.e. they are
separated by about 4\,h. In this month, the average Q2DW-related variance for
both lines is almost on a similar level as the dominating short-term 
intensity variations discussed in section~\ref{sec:shortvar} (ratios of 0.89 
and 0.49 respectively). Hence, a significant fraction of the residual 
variability in January is obviously caused by Q2DW activity. The sharp peak 
at 00:30 LT in July for the relatively faint OH(4-2)P$_1$(14) emission in (f)
is probably caused by measurement uncertainties since such a strong feature at 
this position is not visible in the climatologies of other lines with high 
$N^{\prime}$. Consequently, the less extended LT range with strong short-term
variations in July in (d) compared to the pattern for the entire residual 
variability in (b) can also be explained by this issue. An outlier can
occur more easily in (f) than in (b) of Figure~\ref{fig:di_2lin} since the
residual variances for the required specific time differences are based on
smaller samples. In the case of 00:30 LT in July, there was only a
particularly small number of only 101 data pairs for the calculation of the
variance for $\Delta t = 72$\,h (cf. section~\ref{sec:vartime}), whereas the
sample of 30\,min bins comprised 468 values for this grid point. Moreover, we
estimated the uncertainties in the Q2DW amplitude using half the difference in
the variance $s^2$ for $\Delta t = 24$ and 72\,h (ignoring possible true
systematic differences) as the error for $s^2$ at 24, 48, and 72\,h,
respectively. For those climatological grid points with an amplitude at least
half as strong as the maximum in January, we then derived a satisfying mean
relative uncertainty of about 18\% for both lines in austral summer. On the
other hand, the percentage is about 55\% for OH(4-2)P$_1$(14) in the rest of
the year. 

The location of the maximum values in January agrees well with the 
expected activity period of a westward-propagating Q2DW with a zonal 
wavenumber of 3 \cite<e.g.,>{tunbridge11}. The later peak for OH(4-2)P$_1$(1)
is consistent with the studies of specific time series by \citeA{noll22}. 
Interestingly, the maximum values appear to be more pronounced and more 
clearly separated in the climatologies. The maximum relative amplitudes in 
(e) and (f) are 0.32 and 0.26, respectively. The corresponding values for the 
strong event in 2017 were 0.74 and 0.46. On the other hand, the moderate 
event in 2019 with only a weak LT dependence indicated 0.31 and 0.49, 
respectively. Our Q2DW climatologies should be more representative than the 
two short time intervals that were studied by \citeA{noll22}. However, the 
gaps in the X-shooter data set and the strong year-to-year changes in the 
Q2DW properties \cite<e.g.,>{ern13,gu19,tunbridge11} imply that these 
climatologies are only rough approximations of the long-term averages. For an
illustration of the uncertainties, we recalculated the climatologies without 
the well-covered strong Q2DW event in 2017. The results indicate a decrease 
of the average amplitude in January by 21 and 10\% for the two example lines 
with $N^{\prime} = 1$ and 14. The main decrease is related to the second half
of the night, which showed the highest amplitudes in 2017. As a consequence, 
the morning peak for OH(4-2)P$_1$(1) became more diffuse with two apparent
maxima, whereas the summer pattern in the climatology of the high-$N^{\prime}$
line did not change much. Hence, the eight nights in 2017 had a clear impact
on the Q2DW climatologies in austral summer by increased averages and more
pronounced peaks.

Similar to the properties that were discussed in the previous sections, we 
derived effective Q2DW amplitudes from the nighttime climatologies of the 
whole line set. The results are shown in Figure~\ref{fig:di_eff}b. As 
significant activity of the Q2DW is restricted to the austral summer, the 
typical amplitude relative to the intensity climatology is only of the order
of 10\%. Nevertheless, the dependence of the values on $v^{\prime}$ and
$N^{\prime}$ resembles other plots of this kind. For example, the effective
short-term variations in Figure~\ref{fig:di_eff}a show a good correlation
with $r = +0.77$ despite differences in the outlier distribution. Of course, 
it is also interesting to compare with the corresponding results for the 
maximum amplitude of the Q2DW in 2017. Considering that there are large 
differences in the related samples (10 years vs. eight nights), the 
correlation coefficient of $r = +0.81$ indicates a convincing agreement. 
Moreover, the clear presence of the bump at intermediate $N^{\prime}$ 
supports the conclusion that the mixing of cold and hot populations seems to 
be a general driver for line-specific differences in the strengths of 
perturbations on various time scales.
  
\begin{figure}
\noindent\includegraphics[width=\textwidth]{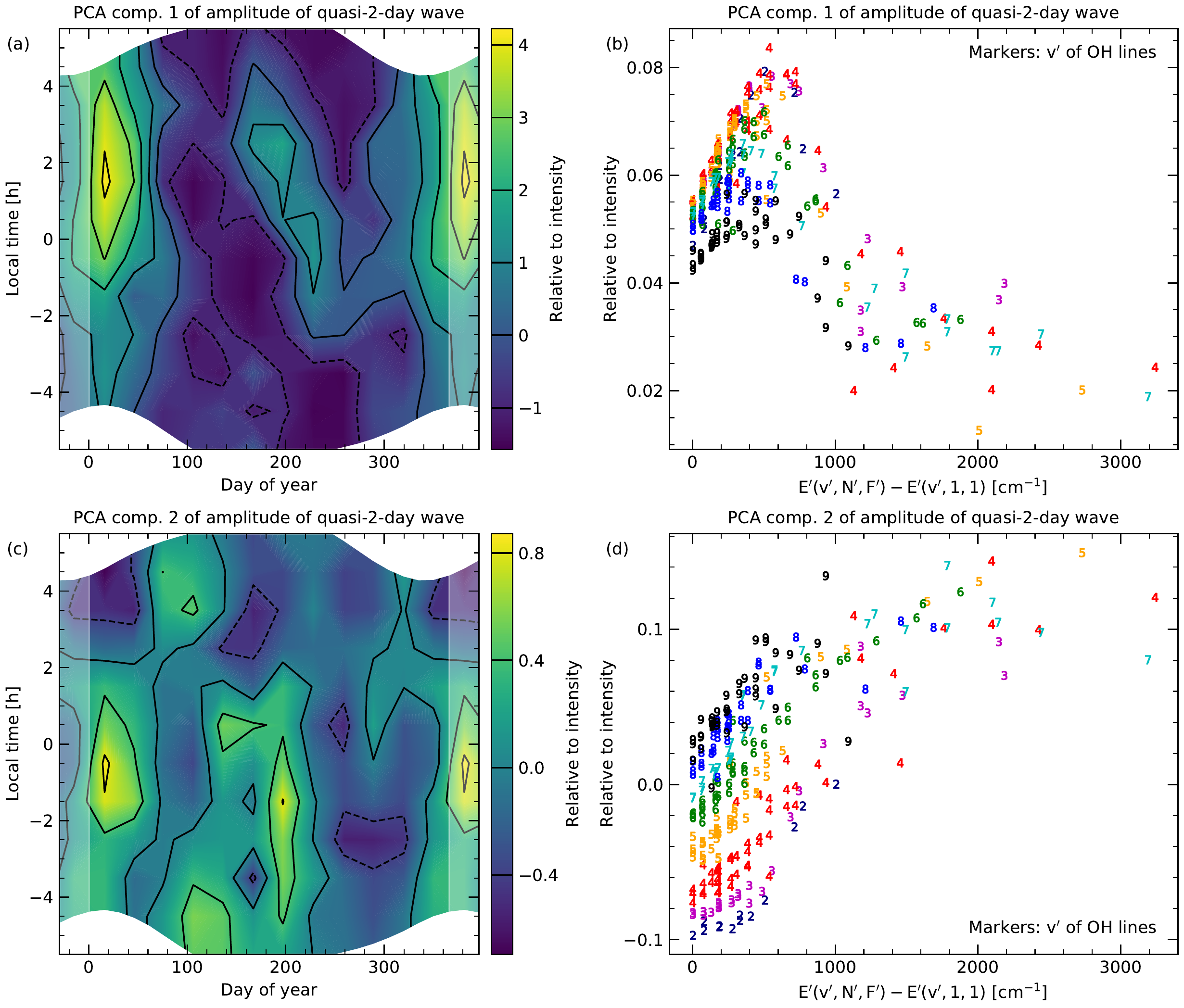}
\caption{Decomposition of climatologies for Q2DW amplitude (examples in 
Figure~\ref{fig:di_2lin}) with principal component analysis for two 
components. The figure is similar to Figure~\ref{fig:sce_pca}.}
\label{fig:aq2dw_pca}
\end{figure}

In Figure~\ref{fig:aq2dw_pca}, we show the results for the PCA of the 
climatologies of the Q2DW amplitude of all 298 lines. As in the case of the
SCE and the short-term variations, we discuss the first two components, which
explain 83.5 and 5.0\% of the variance. The primary basis climatology in (a)
indicates the expected strong maximum in the second half of the night in
austral summer. Based on the full line set, the derived pattern can show
possible features more clearly than the individual climatologies. Hence, there
might be a weak secondary maximum in the same LT range in winter, which would
not contradict other observations \cite<e.g.,>{ern13}. The corresponding
scaling factors in (b) confirm the typical amplitude distribution related to
population mixing. With a correlation coefficient of $+0.95$, the pattern is
very similar to the distribution of the amplitudes for the Q2DW event in 2017.
Apart from comparing the same wave type, the noise-reducing property of the
PCA certainly contributes to this very high $r$. In the same way as for the
SCE and the short-term variations, the second basis climatology in (c) changes
the LTs of the maximum activity. Positive scaling factors in (d) lead to a
shift to earlier LTs. Again, these changes are strongly correlated with the
effective emission height ($r = +0.94$). Hence, Q2DWs also appear to strongly
interact with upward-propagating tides, which confirms the conclusions of
\citeA{noll22} based on the LT dependence of the amplitude of the Q2DW from
2017. Interactions with migrating solar tides were already reported in
previous studies \cite<e.g.,>{hecht10,palo99}.   

The well-localized peaks of the Q2DW amplitude in the individual 
climatologies as shown in the bottom panels in Figure~\ref{fig:di_2lin} allow 
us to quantitatively analyze the relation of height and LT in order to learn
more about the tidal modes that cause the observed activity shifts. For this
purpose, we derived the LT bin in January with the maximum amplitude for the
298 studied OH lines and then averaged the reference emission heights from 
\citeA{noll22} for all lines with the maximum in a given bin. Thereafter, we 
performed a linear regression analysis of the average heights and the central 
times of the LT bins with a length of 1\,h. The relation between height and 
LT turned out to be almost perfectly linear with $r = -0.994$. The slope of
the relation could therefore be derived with high accuracy, amounting to
$-1.23 \pm 0.07$\,km\,h$^{-1}$. This negative phase propagation can be 
converted into the vertical wavelength of a rising wave if the wave period is 
set. For a diurnal tide with a period of 24\,h, the result is 
$29.5 \pm 1.6$\,km, which is remarkably close to the expected wavelength of 
the first symmetric westward propagating mode with a zonal wavenumber of 1.  
\citeA{forbes95} provide 27.9\,km based on calculations. This migrating solar
tide, which is also known as DW1, is the dominating tidal mode at low 
latitudes \cite<e.g.,>{smith12}. The most important semidiurnal mode, the 
westward-migrating SW2 tide with a zonal wavenumber of 2 would need a 
wavelength of about 15\,km in order to fit. However, its wavelength is 
clearly longer than in the case of the DW1 \cite{smith12}. Hence, it is 
likely that the line-dependent LT shift of the maximum Q2DW activity in 
January can mostly be explained by the interaction of Q2DW and DW1. The 
significant contribution of the Q2DW in 2017 to the climatologies seems to be
important for this result since the LT difference for the two example lines 
as reported above would otherwise be more uncertain. The Q2DW in 2017 had a 
most probable wavelength of about 32\,km during the analyzed eight nights 
\cite{noll22}. As this is comparable to the DW1, strong interactions are 
likely. This statement is supported by the fact that the Q2DW from 2019, 
which did not show a strong LT dependence, had a very long vertical 
wavelength.       

The interpretation of the line-dependent shifts of the patterns in the 
climatologies of the SCE and short-term variations is more difficult as the
most striking features are broader and are partly cut by the nighttime 
limits. Nevertheless, there are clear indications that the situation for the 
Q2DW in austral summer can be generalized. The best examples are probably the 
climatologies of the residual variability for the two reference lines in the
top row of Figure~\ref{fig:di_2lin}. These climatologies include the Q2DW as 
well as the GW activity. In these cases, the variability in January and July
show a similar dependence on LT, at least with respect to the main peaks. The
LT difference for the maximum in July is probably about 4\,h with an 
uncertainty of 1\,h, i.e. very similar to the result for the Q2DW in January 
shown in the bottom panels. The extracted short-term variations in panels (c)
and (d) of Figure~\ref{fig:di_2lin} are also consistent with this estimate. In
the case of the SCE shown in Figure~\ref{fig:sce_2lin}, the interpretation is
most difficult but the minimum difference should be 2 to 3\,h. Hence, the DW1
is also the most likely tidal mode for the origin of the line-dependent
changes in the climatological patterns of the short-term variations (i.e.
especially GW activity) and the SCE. Nevertheless, the OH intensity variations
are also affected by other tidal modes (with longer vertical wavelengths) as
the analysis of the intensity climatologies in section~\ref{sec:intensity} 
indicates. There, the location of the tidal features appears to be 
relatively robust with respect to the line parameters.

\section{Conclusions}\label{sec:conclusions}

We studied intensity time series (binned in 30\,min steps) of 298 OH lines 
with respect to climatological variability patterns based on almost 90,000 
near-infrared spectra taken at Cerro Paranal in Chile in the time interval 
from October 2009 to September 2019. Different 2D climatologies were 
calculated for local time (LT) and day of year with grid steps and (minimum) 
data selection radii of 1\,h and 1 month, respectively. 

The climatologies of the OH intensities relative to the mean revealed a 
strong dependence on line parameters such as the upper vibrational level 
$v^{\prime}$ and upper rotational level $N^{\prime}$. Using nonnegative 
matrix factorization for the decomposition of the observed patterns, we could
clearly separate two major components. First, there is a relatively stable 
variability structure which can be explained by tidal features, which are 
strongest in the middle of the year. Similar features can also be observed in 
the temperature and atomic oxygen concentration at the altitudes with the 
strongest OH emission. Their amplitude seems to maximize for intermediate 
rotational energies between 400 to 800\,cm$^{-1}$. A similar amplitude
distribution was previously discovered by \citeA{noll22} for a quasi-2-day
wave (Q2DW) which was investigated in a time interval of eight nights in 2017
based on the same X-shooter data set. Another Q2DW in 2019 (seven nights) also
showed this feature. It can obviously be explained by assuming that the
population distribution for each $v^{\prime}$ can be described by the
combination of a cold thermalized population with the effective ambient
temperature at the OH emission heights and a hot nonthermalized population
with $v^{\prime}$-dependent pseudo temperatures \cite{noll20}. The maximum
tidal variations are then found where both populations show similar
contributions on average since the steep decrease of the cold population with
increasing $N^{\prime}$ there causes a particular high variability of the
population mixing if the ambient temperature changes. The second component of
the relative intensity climatologies is characterized by a general decrease
from the evening to the morning with amplitude maxima near the equinoxes.
Assuming an exponential function, the effective time constant of the decrease
is $3.3 \pm 0.2$\,h with possible seasonal variations. The contribution to the 
combined climatologies strongly decreases with increasing $v^{\prime}$ and 
$N^{\prime}$ and vanishes almost completely for the highest rotational 
levels. This behavior resulted in a strong anticorrelation with the 
effective line emission heights based on phase measurements in the X-shooter
and related height-resolved SABER data for the Q2DW event in 2017 
\cite{noll22}. Supported by the results of \citeA{marsh06}, this component 
can be explained by the particularly strong decay of the nighttime population
of atomic oxygen, which is essential for the OH production and is mostly 
produced by photolysis of molecular oxygen at daytime, at the lowest OH 
emission altitudes below 84\,km. 

We also calculated climatologies of the solar cycle effect (SCE) relative to 
the corresponding intensity climatologies for 27-day averages of the solar 
radio flux. The effective SCE values derived from the entire nighttime 
climatologies show a large range between about 8 and 23\% per 100 solar flux 
units (sfu), which was not observed before but is consistent with previous 
results if it is considered which OH lines contributed to the analysis. The 
lowest SCE values were found for the lines with the lowest $v^{\prime}$ and 
$N^{\prime}$. Between intermediate and high rotational energies, no clear 
trend was seen. This distribution can be explained by the striking structure 
of the SCE climatologies with values between slightly negative and more than 
$+50$\% per 100\,sfu and its change depending on the line parameters. A 
principal component analysis (PCA) revealed that the primary component is 
characterized by a strongly positive effect in the second part of the night 
around July and only weak effects otherwise. This component also shows the 
highest values for intermediate $N^{\prime}$, i.e. differences in the 
sensitivity of the OH level populations to changes in the ambient temperature
is the main reason for the observed discrepancies. The second component 
essentially describes a shift of this pattern in LT direction. As indicated 
by a very high correlation coefficient of $+0.96$, the LT of the maximum is 
later for lines with lower effective emission height. This effect obviously 
contributes to the low effective nighttime SCEs for lines with low 
$v^{\prime}$ and $N^{\prime}$ as the extension of the maximum feature in LT 
direction appears to be reduced by the end of the night. The shift of the SCE
pattern can be best understood in terms of the impact of upward-propagating 
perturbations, which seem to change the sensitivity of the OH emission to 
atmospheric effects of solar activity such as higher atomic oxygen production
and higher temperatures. As we investigate climatologies, the perturbations 
are most likely related to solar tides.    

By correcting the OH intensity data for the mean climatologies and the SCE
using the corresponding solar radio flux, we could also study climatologies
of the residual variability consisting of the standard deviations for the
selected subsamples for each climatological grid point. The results show 
maximum values around the solstices. With the derivation of the mean variance 
as a function of the time difference of data pairs for each subsample, we 
were able to distinguish between different variability sources. Time scales
up to several hours are most important. Such short-term variations show a
clear maximum in June and July, which is probably related to gravity waves
(GWs) that tended to be generated in the south towards the Andean winter hot 
spot and reached Cerro Paranal by favorable propagation conditions either 
directly, via wave ducts, or as secondary waves. These waves may also play an
important role for the strong SCE effect that is present in a similar region
of the climatologies. With respect to the activity maximum in austral winter,
short-term variations only explain a part of the residual variability as
especially January is characterized by a significant contribution of Q2DWs,
which we also measured. The remaining activity on short time scales might 
mainly be related to GWs that originate from deep convection in the north and 
east on the other side of the Andes. For the climatologies of the short-term 
variations and the Q2DW amplitude, we also calculated effective values for 
each OH line. Compared to the SCE, they are less affected by the nighttime 
limitations as the LTs with the highest activity were not close to the 
twilight for all lines. As a consequence, the line-dependent distributions of 
both properties showed a clearer bump at intermediate $N^{\prime}$. As the
first PCA components also revealed this feature, it seems that the 
line-dependent mixing of cold and hot populations has a major impact on the 
amplitude of various kinds of variations with time scales that can differ by 
several orders of magnitude. Similar to the SCE, the second PCA components
indicate clear shifts of the climatological patterns in LT direction that
are strongly correlated with the effective emission heights of the lines.
Hence, the impact of tides on the sensitivity of OH emission to 
perturbations can also be generalized. In order to learn more about the
relevant tidal modes, we used the well-defined LTs of the maximum Q2DW 
amplitude in January (with clear impact of the strong event in 2017) of all
298 lines in order to link them to the corresponding effective emission 
heights. The resulting height--LT relation is almost perfectly linear with a 
slope of $-1.23 \pm 0.07$\,km\,h$^{-1}$ that can be best explained by a wave 
period of 24\,h and a vertical wavelength of about 30\,km. As the 
climatologies for short-term variations and SCE show similar pattern shifts, 
it appears that OH-based studies of GWs, Q2DWs, solar activity, and possibly 
other variability sources at Cerro Paranal and similar locations are 
significantly affected by the westward-propagating diurnal tide with zonal 
wavenumber 1, DW1. This tidal mode can act by direct interactions with the 
other perturbations and/or indirectly via the change of the OH-related 
chemistry, which particularly depends on the atomic oxygen profile.

\section{Open Research}
The basic X-shooter NIR-arm data for this project originate from the ESO 
Science Archive Facility at \mbox{http://archive.eso.org} (open access for all
data used) and are related to various observing programs that were carried 
out between October 2009 and September 2019. The raw spectra were processed 
(using the corresponding calibration data) and then analyzed.
Data of the analysis of eight nights in 2017 and seven nights in 2019 with 
respect to specific Q2DWs were already published \cite{noll22ds}. We used
the resulting line-dependent wave amplitudes and effective emission heights. 
Concerning this study, the line-specific time series (binned in 30\,min 
steps) for the calculation of the climatologies and the results of the 
analysis as partly shown in the figures can be obtained from the public
repository Zenodo at \mbox{http://zenodo.org/record/7826060} \cite{noll23ds}.

%AGU requires an Availability Statement for the underlying data needed to understand, evaluate, and build upon the reported research at the time of peer review and publication.

%Authors should include an Availability Statement for the software that has a significant impact on the research. Details and templates are in the Availability Statement section of the Data and Software for Authors Guidance: \url{https://www.agu.org/Publish-with-AGU/Publish/Author-Resources/Data-and-Software-for-Authors#availability}

%It is important to cite individual datasets in this section and, and they must be included in your bibliography. Please use the type field in your bibtex file to specify the type of data cited. Options include [Dataset], [Software], [ComputationalNotebook], [Collection].
%Example:
%
%@misc{https://doi.org/10.7283/633e-1497,
%  doi = {10.7283/633E-1497},
%  url = {https://www.unavco.org/data/doi/10.7283/633E-1497},
%  author = {de Zeeuw-van Dalfsen, Elske and Sleeman, Reinoud},
%  title = {KNMI Dutch Antilles GPS Network - SAB1-St_Johns_Saba_NA P.S.},
%  publisher = {UNAVCO, Inc.},
%  year = {2019},
%  type = {dataset}
%}

%For physical samples, use the IGSN persistent identifier, see the International Geo Sample Numbers section:
%\url{https://www.agu.org/Publish-with-AGU/Publish/Author-Resources/Data-and-Software-for-Authors#IGSN}
%%%%%%%%%%%%%%%%%%%%%%%%%%%%%%%%%%%%%%%%%%%%%%%

\acknowledgments
Stefan Noll is financed by the project NO\,1328/1-3 of the German Research
Foundation (DFG). The authors thank Sabine M\"ohler from ESO for her support
with respect to the X-shooter calibration data. Moreover, the authors are
grateful to the two anonymous reviewers for their valuable comments. 

%This section is optional. Include any Acknowledgments here.
%The acknowledgments should list:\\
%All funding sources related to this work from all authors\\
%Any real or perceived financial conflicts of interests for any author\\
%Other affiliations for any author \newline that may be perceived as having a conflict of interest with respect to the results of this paper.\\
%\clearpage
%It is also the appropriate place to thank colleagues and other contributors. AGU does not normally allow dedications.\\
%\input{methods}
%Hello here is some text in Nick's file
%\input{data}

%% ------------------------------------------------------------------------ %%
%% References and Citations

%%%%%%%%%%%%%%%%%%%%%%%%%%%%%%%%%%%%%%%%%%%%%%%
%
% \bibliography{<name of your .bib file>} don't specify the file extension
%
% don't specify bibliographystyle

% In the References section, cite the data/software described in the Availability Statement (this includes primary and processed data used for your research). For details on data/software citation as well as examples, see the Data & Software Citation section of the Data & Software for Authors guidance
% https://www.agu.org/Publish-with-AGU/Publish/Author-Resources/Data-and-Software-for-Authors#citation

%%%%%%%%%%%%%%%%%%%%%%%%%%%%%%%%%%%%%%%%%%%%%%%

\bibliography{Nolletal2023a}

%Reference citation instructions and examples:
%
% Please use ONLY \cite and \citeA for reference citations.
% \cite for parenthetical references
% ...as shown in recent studies (Simpson et al., 2019)
% \citeA for in-text citations
% ...Simpson et al. (2019) have shown...
%
%
%...as shown by \citeA{jskilby}.
%...as shown by \citeA{lewin76}, \citeA{carson86}, \citeA{bartoldy02}, and \citeA{rinaldi03}.
%...has been shown \cite{jskilbye}.
%...has been shown \cite{lewin76,carson86,bartoldy02,rinaldi03}.
%... \cite <i.e.>[]{lewin76,carson86,bartoldy02,rinaldi03}.
%...has been shown by \cite <e.g.,>[and others]{lewin76}.
%
% apacite uses < > for prenotes and [ ] for postnotes
% DO NOT use other cite commands (e.g., \citet, \citep, \citeyear, \citealp, etc.).
% \nocite is okay to use to add references from your Supporting Information
%

\end{document}